\documentclass[acmsmall,screen]{acmart}

\setcopyright{acmlicensed}
\copyrightyear{2026}
\acmYear{2026}
\acmDOI{XXXXXXX.XXXXXXX}
\acmConference[FSE '26]{The ACM International Conference on the Foundations of Software Engineering}{July 05--09,
  2026}{Montreal, Canada}

\acmISBN{978-1-4503-XXXX-X/2018/06}

\usepackage[linesnumbered,ruled,vlined]{algorithm2e}
\usepackage{tabularx}
\usepackage{adjustbox}
\usepackage{booktabs}
\usepackage{threeparttable}
\usepackage{makecell}
\usepackage{enumitem}
\usepackage{multirow}
\usepackage{colortbl}
\usepackage{pifont}
\usepackage{caption}
\newcommand{\revise}[1]{#1} 

\usepackage[table]{xcolor}
\usepackage[svgnames]{xcolor}
\definecolor{myc2}{HTML}{3596b5} 
\definecolor{myc}{HTML}{adc5cf} 
\usepackage[most]{tcolorbox}
\newtcolorbox{mybox}{
  colback=gray!10,     
  colframe=black!75,   
  boxrule=0.5pt,       
  arc=2pt,             
  outer arc=2pt,
  left=6pt,            
  right=6pt,           
  top=6pt,             
  bottom=6pt,          
  boxsep=0pt,          
}
\begin{document}

\title{TypePro: Boosting LLM-Based Type Inference via Inter-Procedural Slicing}

\author{Teyu Lin}
\affiliation{
  \institution{Xiamen University}
  \city{Xiamen}
  \state{Fujian}
  \country{China}
}
\email{linteyu@stu.xmu.edu.cn}

\author{Minghao Fan}
\affiliation{
  \institution{Xiamen University}
  \city{Xiamen}
  \state{Fujian}
  \country{China}
}
\email{fanminghao@stu.xmu.edu.cn}

\author{Huaxun Huang}
\affiliation{
  \institution{Xiamen University}
  \city{Xiamen}
  \state{Fujian}
  \country{China}
}
\email{huanghuaxun@xmu.edu.cn}

\author{Zhirong Shen}
\affiliation{
  \institution{Xiamen University}
  \city{Xiamen}
  \state{Fujian}
  \country{China}
}
\email{shenzr@xmu.edu.cn}

\author{Rongxin Wu}
\affiliation{
  \institution{Xiamen University}
  \city{Xiamen}
  \state{Fujian}
  \country{China}
}
\email{wurongxin@xmu.edu.cn}

\begin{abstract}
Dynamic languages (such as Python and JavaScript) offer flexibility and simplified type handling for programming, but this can also lead to an increase in type-related errors and additional overhead for compile-time type inference. As a result, type inference for dynamic languages has become a popular research area. Existing approaches typically achieve type inference through static analysis, machine learning, or large language models (LLMs). However, current work only focuses on the direct dependencies of variables related to type inference as the context, resulting in incomplete contextual information and thus affecting the accuracy of type inference. 
To address this issue, this paper proposes a method called TypePro, which leverages LLMs for type inference in dynamic languages. TypePro supplements contextual information by conducting inter-procedural code slicing. Then, TypePro proposes a set of candidate complex types based on the structural information of data types implied in the slices, thereby addressing the lack of domain knowledge of LLMs.
We conducted experiments on the ManyTypes4Py and ManyTypes4TypeScript datasets, achieving Top-1 exact match (EM) rates of 88.9\% and 86.6\%, respectively. 
Notably, TypePro improves the Top-1 Exact Match by 7.1 and 10.3 percentage points over the second-best approach, showing the effectiveness and robustness of TypePro.

\end{abstract}




\maketitle

\section{Introduction}

Dynamic languages like Python and JavaScript do not require explicit type declarations for variables, meaning a variable’s type is determined only at runtime. This enables faster development and greater flexibility in how variables are used, but it also introduces issues such as type errors that only surface at runtime and additional runtime overhead for type-related inference~\cite{Typilus}. To address issues caused by dynamic typing, Python introduced a standard for type annotations in PEP 484~\cite{pep484}. The JavaScript superset TypeScript ~\cite{bierman2014understanding} provides static type checking at compile time and is then compiled (transpiled) to plain JavaScript for execution at runtime~\cite{hellendoorn2018deep}. These solutions likewise aim to mitigate problems introduced by dynamic typing. However, writing type annotations manually is both time-consuming and error-prone~\cite{DoMachineLearning}, which has motivated research into type inference for dynamic languages.

Existing research has explored dynamic type inference from several aspects. First, industry-used automated type inference tools (such as \text{mypy}~\cite{mypy}) mainly rely on pre-defined typing rules. However, such pre-defined rules are often incomplete, which has limited the inference capability of these tools. Second, classification-based methods have cast type inference as a multi-class classification task and have use machine learning models to automatically predict the type of a code fragment—for example, DeepTyper~\cite{hellendoorn2018deep} for JavaScript, DLInfer for Python~\cite{yan2023dlinfer}, and Type4Py ~\cite{mir2022type4py}. These methods require substantial amounts of labeled data for training and are heavily dependent on the quality and comprehensiveness of the training set. Consequently, their prediction accuracy often declines when encountering types not included in the training data, such as user-defined types. In recent years, with the advancement of large language models (LLMs), a class of generative LLM-based approaches has emerged—e.g., TypeGen~\cite{peng2023generative} and Tiger~\cite{wang2024tiger}. These methods have improved type inference by combining prompt engineering with program-analysis-provided context (such as program slices and suggestions of likely third-party libraries or user-defined types) and other feedback mechanisms.

\begin{figure}[t]
    \centering
    \includegraphics[width=1\textwidth]{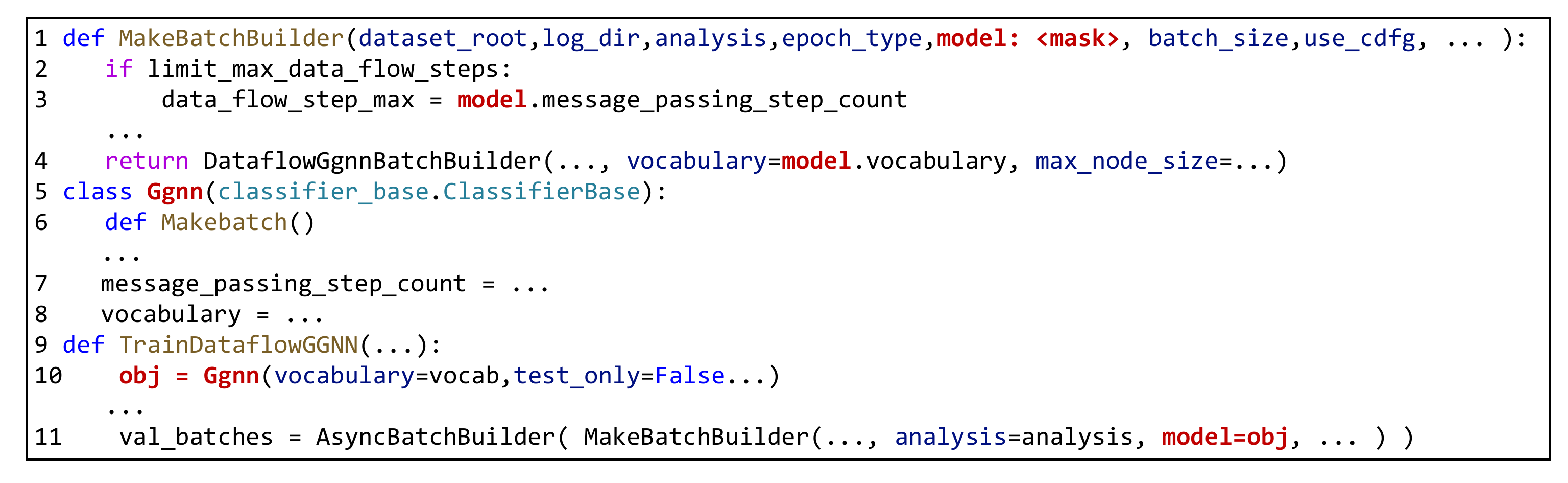}
    \caption{The motivating example where \texttt{model} serves as the target variable for data type inference.}
    \Description{}
    \label{fig:example code}
\end{figure}

Although existing approaches have made some progress, a major limitation is that the program context they rely on is often incomplete. Type inference typically requires diverse contextual information: variables may be passed to and modified by different functions, functions may be invoked from other files, and variables may depend on global state or external inputs. Acquiring such context requires inter-procedural data-flow analysis. For example, to infer the type of the parameter \texttt{model} in line 1 of Figure~\ref{fig:example code}, one must perform inter-procedural backward data-flow analysis on the function \texttt{TrainDataflowGGNN} (lines 9–11). From line 10 it can be determined that the variable is of the user-defined type \texttt{Ggnn} (the class definition of \texttt{Ggnn} appears at line 5). Because they lack inter-procedural data-flow analysis, existing methods (e.g., \cite{mir2022type4py, peng2023generative, wang2024tiger}) cannot accurately infer the type of \texttt{model}.

Based on the above observation, we propose TypePro, an LLM-based type inference method for dynamic languages. Specifically, with a knowledge base of third-party libraries and the project's own data types, TypePro performs inter-procedural code slicing to extract contextual information of the target variable. To address the issue of LLMs lacking knowledge of third-party libraries and custom types, TypePro selects a candidate list of data types from this knowledge base whose structural information is similar to that implied in the slice. These two pieces of information are then combined into a prompt, guiding the LLM to generate data types.

We evaluated TypePro in the widely used ManyTypes4Py ~\cite{mir2021manytypes4py} and ManyTypes4TypeScript ~\cite{Manytypes4typescript} datasets. The evaluation results show that TypePro achieved Top-1 accuracies of 87.8\% (ManyTypes4Py) and 86.6\% (ManyTypes4TypeScript). Notably, TypePro improves the Top-1 Exact Match by 7.1 and 10.3 percentage points over the second-best
approaches in ManyTypes4Py and ManyTypes4TypeScript, showing the effectiveness and robustness of TypePro.

In summary, this paper makes the following contributions:

\begin{itemize}[leftmargin=*]
\item To the best of our knowledge, we propose the first inter-procedural static code-slicing method for type inference of dynamic programming languages. 

\item We design a new candidate data type selection strategy, which uses the structural information of the data type inferred from the slice to perform type inference.

\item We implemented an automated tool called TypePro and performd extensive experiments against baselines. The results show that TypePro is effective in both Python and TypeScript, and it outperforms existing work in Top-1 accuracy.
\end{itemize}
\section{Related Work}
In this section, we categorize existing type inference approaches into three main types: rule-based methods, classification-based methods, and generation-based methods. Each type is discussed in detail below.

\subsection{Rule-Based Methods}
These methods relied on a predefined set of type rules and employed static analysis to explore variables in the code. When a target variable met a specific pre-defined rule, its type was inferred accordingly~\cite{howFar}. Examples of such methods included Pyright and Pylance~\cite{microsoft_pyright} released by Microsoft, pyType released by Google~\cite{google_pytype}, and the official Python type checker mypy~\cite{mypy}. In the JavaScript/TypeScript ecosystem, relevant approaches included the official TypeScript compiler tsc and Babel. Additionally, several cross-language methods existed~\cite{xu2016python, DataFlowInference, Symbolicabstract}.

These rule-based approaches were known for their high precision, enabling accurate type inferences for the scenarios they encompassed. Nonetheless, a primary challenge associated with these methods was the considerable initial effort needed to establish a comprehensive set of rules, given the immense workload involved. Moreover, a significant limitation of these methods was their insufficient coverage when addressing external function calls and the dynamic features of programming languages. Without developing new rules tailored specifically to these situations, achieving satisfactory results remained challenging.

\subsection{Classification-Based Methods}
These methods frame type inference as a machine learning classification task, typically employing supervised learning with features extracted from code and target types represented as vectors. In the TypeScript ecosystem, DeepTyper~\cite{hellendoorn2018deep} frames type inference as a machine translation problem using a bidirectional RNN to encode code context. LambdaNet~\cite{wei2020lambdanetprobabilistictypeinference} adopts a graph neural network on type dependency graphs to emulate static analysis.

In Python, DLInfer~\cite{yan2023dlinfer} and Type4Py~\cite{mir2022type4py} combine static code slicing with neural models. Type4Py constructs a large vocabulary and maps it to type clusters in a high-dimensional space via a hierarchical neural network.

These machine learning methods benefit from extensive prior research and achieve competitive performance on benchmarks. However, they are constrained by a fixed label vocabulary, which limits generalization to unseen types~\cite{peng2023generative}. Performance also heavily depends on high-quality training data, and insufficient or noisy data can degrade effectiveness. Additionally, they often lack interpretability, making it difficult to understand the rationale behind predictions.

\subsection{Generation-Based Methods}
Generative models are designed to generate missing information, making them suitable for tasks like code generation and infilling. Typically, these approaches place a mask at the location requiring a type annotation and task the model with filling in the type. Common models for such tasks include CodeT5~\cite{codeT5}, UnixCoder~\cite{guo2022unixcoderunifiedcrossmodalpretraining}, and InCoder~\cite{fried2023incodergenerativemodelcode}, which are pre-trained on large-scale code corpora to predict subsequent code sequences given context. During inference, decoding strategies are used to generate candidate code.

Compared to classification-based methods, generative models often achieve better performance across a wider range of scenarios. However, their effectiveness remains limited. Some works, such as TIGER~\cite{wang2024tiger}, enhance generative models by incorporating similarity computation for type recommendation, improving their ability to handle user-defined types. Nevertheless, these methods—whether directly using base models like CodeT5 or InCoder, or enhanced frameworks like TIGER—rely heavily on incomplete information, typically restricted to the immediate local context. This often omits critical information and limits the ability to infer user-defined types. Even TIGER, despite its similarity-based recommendation, lacks essential structural information about related types and inter-procedural context.

Large language models (LLMs) have emerged as a prominent research topic in recent years. Trained on large corpora of text and code and comprising billions of parameters, these models demonstrate human-like capabilities in processing and generating text and can be viewed as a class of generative models. Notably, LLMs have been applied to type-inference tasks.
Similar to the direct use of generative models, methods based on LLMs may employ a direct question-answering approach, such as TypeGen~\cite{peng2023generative}. Unlike approaches that require training a dedicated generative model from scratch, LLM-based methods leverage general-purpose pre-trained models, significantly reducing development time. Additionally, these methods often incorporate techniques like chain-of-thought prompting to enhance reasoning and improve performance. For example, TypeGen~\cite{peng2023generative} combines LLMs with static analysis by providing code slices as input to the LLMs and translating static analysis results into a chain-of-thought format, i.e., expressing static analysis outcomes in natural language. This approach has demonstrated strong performance on type inference benchmarks. Nevertheless, as previously mentioned, it still faces the same limitations in code slicing and handling user-defined types.

\section{Motivation}
\begin{table}[t]
\renewcommand{\arraystretch}{1} %
    \centering
        \caption{The results of TypePro and baselines for the example in Figure ~\ref{fig:example code}}
    \begin{tabular}{|l|l|l|}
    \cline{1-3}
        \textbf{Approach} &\textbf{Baseline} &\textbf{Inferred Data Type}\\
        \cline{1-3}
        GEN & TypeGen & \makecell[l]{1. \texttt{Typing.Type} \\ 2. \texttt{model.vocabulary} \\ 3. \texttt{model}} \\
        \cline{1-3}
        CLS & Hityper & \makecell[l]{1. \texttt{int} \\ 2. \texttt{str} \\}\\
        \cline{1-3}
        \multirow{2}{*}{GEN+SIM} & Tiger & \makecell[l]{1. \texttt{M} \\2. \texttt{DataflowGgnnBatchBuilder} \\3. \texttt{Ggnn}}\\
        \cline{2-3}
        &TypePro&\makecell[l]{\texttt{Ggnn}}\\
    \cline{1-3}
    \end{tabular}
    \label{tab:example results}
\end{table}
\begin{figure}[t]
    \centering
    \includegraphics[width=1\textwidth]{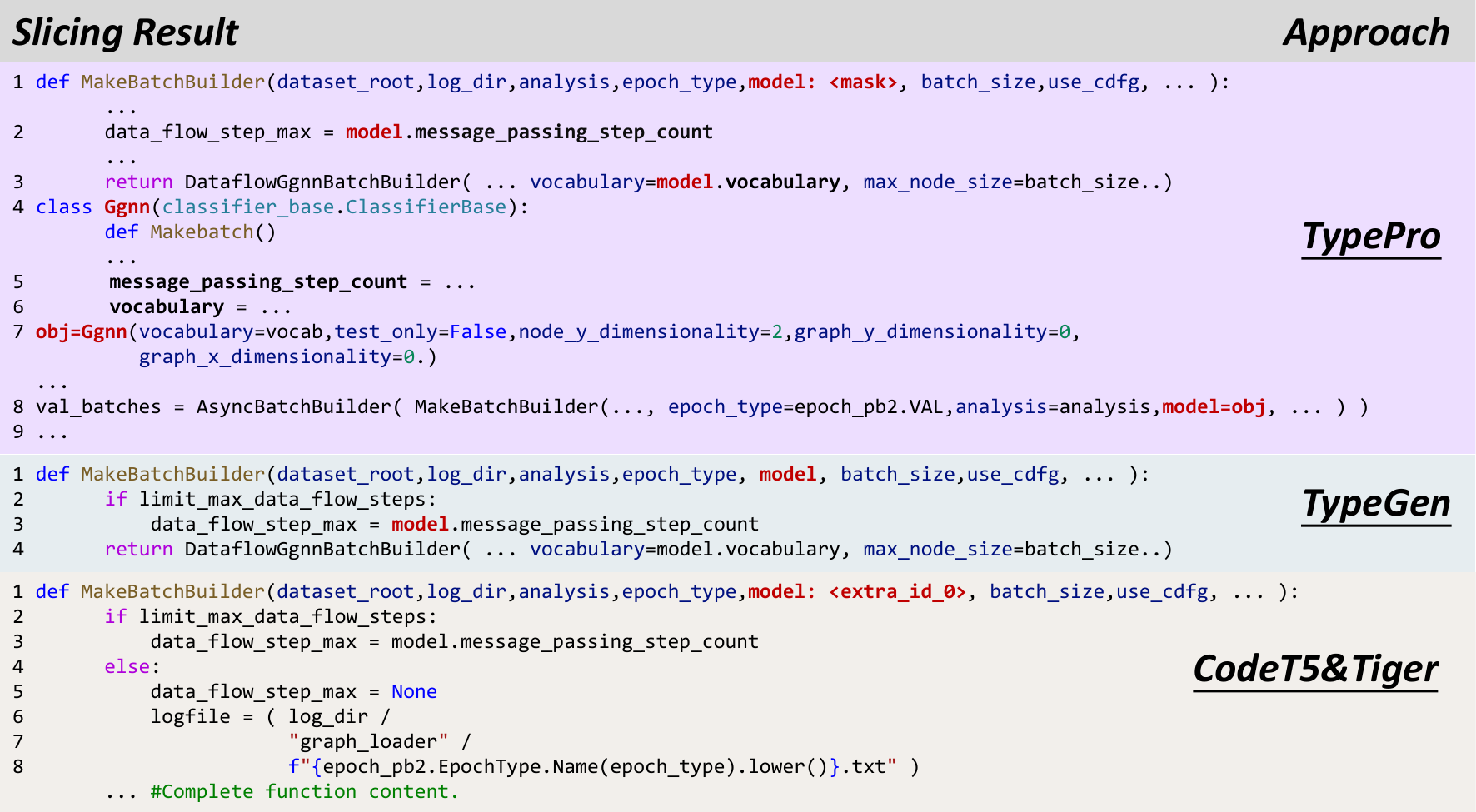}
    \caption{The slices of Figure ~\ref{fig:example code} for TypePro, TypeGen and Tiger.}
    \Description{}
    \label{fig:slicing result}
\end{figure}
\label{sec:motivation}

In this section, we evaluate state-of-the-art approaches to identify their limitations in type inference. We apply HiTyper~\cite{Hityper}, TypeGen~\cite{peng2023generative}, and TIGER~\cite{wang2024tiger} to the example shown in Figure~\ref{fig:example code}. Specifically, HiTyper is a classification-based (CLS) method that constructs a Type Dependency Graph (TDG) to represent the composition and dependencies of variables, and then performs type inference using a similarity-based model. TypeGen is a generation-based (GEN) approach that slices code using TDG and integrates the natural language summary of static analysis with type inference examples to construct a chain of thought prompt and generate types. TIGER is considered as a generation-and-similarity-based (GEN+SIM) approach, employing a pre-trained generative model along with a similarity computation model to rank possible data types. It is worth noting that there are other CLS and GEN approaches available, while we selected one representative state-of-the-art method from each category for illustration.

Table~\ref{tab:example results} presents the output of TypePro and baseline methods in the example illustrated in Figure ~\ref{fig:example code}. We now discuss their results according to the categories of different methods as follows.

\begin{itemize}[leftmargin=*]
    \item \textbf{CLS Approach.} It can be observed that the current state-of-the-art classification-based type inference technique, HiTyper, fails to accurately infer the specific type of the parameter \texttt{model}. The root cause lies in the fact that the deep learning model of HiTyper relies heavily on large amounts of annotated data for training, making its performance highly dependent on the quality and coverage of the training set. Therefore,  the success rate of HiTyper can drop when HiTyper encounters types not present in the training data (e.g., user-defined types). Specifically, for the example in Figure~\ref{fig:example code}, HiTyper incorrectly output primitive types such as \texttt{int} and \texttt{str}, therefore failing to recognize that the variable should be in the user-defined type called \texttt{Ggnn}.
    \item \textbf{GEN Approach.} It can be observed that the current state-of-the-art generation-based approach, TypeGen, also fails to infer the type of the parameter \texttt{model}. The main reason is that the slices generated by TypeGen only include the data flow of parameters within the function, lacking any data flow information on the actual arguments of the parameters. For instance, when the TypeGen method performs code slicing on the code shown in Figure~\ref{fig:example code}, it only considers the code related to the parameter model inside the function \texttt{MakeBatchBuilder}, as well as the statements calling the \texttt{MakeBatchBuilder} function (as shown in Figure~\ref{fig:slicing result}). These sliced segments can only suggest that the parameter \texttt{model} might be of the same type as the actual argument \texttt{obj}. However, since the slice does not include the type information of \texttt{obj}, it is difficult to further infer the concrete type of the parameter \texttt{model}.
    
    \item \textbf{GEN+SIM Approach.} Although TIGER covers user-defined types and third-party library types and is capable of inferring the correct data type for \texttt{model}, it only ranks the correct data type in the third position. This is primarily because the type inference process of TIGER relies solely on the local context surrounding the target location (as shown in Figure \ref{fig:slicing result}) and ranks candidate data types through generative likelihood computation and the extracted contextual information. However, depending exclusively on the immediate context before and after the target variable makes it difficult to fully reconstruct the structure of non-primitive types, especially when the definitions of these types might be distributed across different locations within the project.
\end{itemize}

In the scenario depicted in Figure \ref{fig:example code}, to let LLM accurately determine the type of the parameter \texttt{model} in the function \texttt{MakeBatchBuilder}, an automated tool should: (1) analyze the call statements related to \texttt{MakeBatchBuilder} (i.e., lines 9–11), (2) perform code slicing on the actual argument \texttt{obj} passed to \texttt{MakeBatchBuilder}, and (3) reconstruct the structural features of potential user-defined data types (e.g., \texttt{model.message\_passing\_step\_count}, as shown in line 3 of Figure \ref{fig:example code}). Specifically, the def statement of \texttt{obj} occurs at line 10, and its usage is located at line 11. Including these statements in the code slice helps infer the type of the actual argument \texttt{obj}. It is worth noting that the definition of \texttt{obj} is accomplished through a call to \texttt{Ggnn}. Therefore, it can be inferred that the type of \texttt{obj} is either consistent with the return type of \texttt{Ggnn} or corresponds to the \texttt{Ggnn} class itself.

To further infer the type of \texttt{obj}, one can perform backward slicing starting from \texttt{Ggnn}, incorporating its function/class definition (e.g., the signature at line 5) into the code slice. Finally, by analyzing the transitions from the target function parameter \texttt{model} to the argument \texttt{obj}, and further to the return type of the function \texttt{Ggnn}, together with the structural information (e.g., \texttt{model.message\_passing\_step\_count}), TypePro can guide LLM to determine that the type of \texttt{model} is \texttt{Ggnn}.

\revise{To summarize, the key contributions of TypePro comparing to existing approaches are as follows:}
\begin{itemize}[leftmargin=*]
\item \revise{TypePro incorporates inter-procedural analysis of function calls for type inference, in contrast to TypeGen's code slicing that is limited to intra-procedural analysis. For example, the slicing process of TypePro includes \texttt{model} $\rightarrow$ \texttt{obj} flow that depicts the call of \texttt{MakeBatchBuilder} in line 11 of Figure~\ref{fig:example code}. Compared to Tiger, which extracts code snippets around the target variable based on heuristics, the code slicing process of TypePro can provide context information that is data-flow related to the target variable, and this contextual information has been experimentally shown to improve the type inference precision (see Section 4.3 for details).}


\item \revise{Through this slicing process, TypePro restores the structural information of user-defined data types (e.g., the \texttt{Ggnn} type containing \texttt{message\_passing\_step\_count} and \texttt{vocabulary} fields in lines 2-3 of Figure ~\ref{fig:example code}) and suggests candidate types matching this structure to the LLM, thereby improving effectiveness (see Sections 4.2 and 4.4).}

\end{itemize}
\section{Approach}
\begin{figure}[t]
    \centering
    \includegraphics[width=1\textwidth]{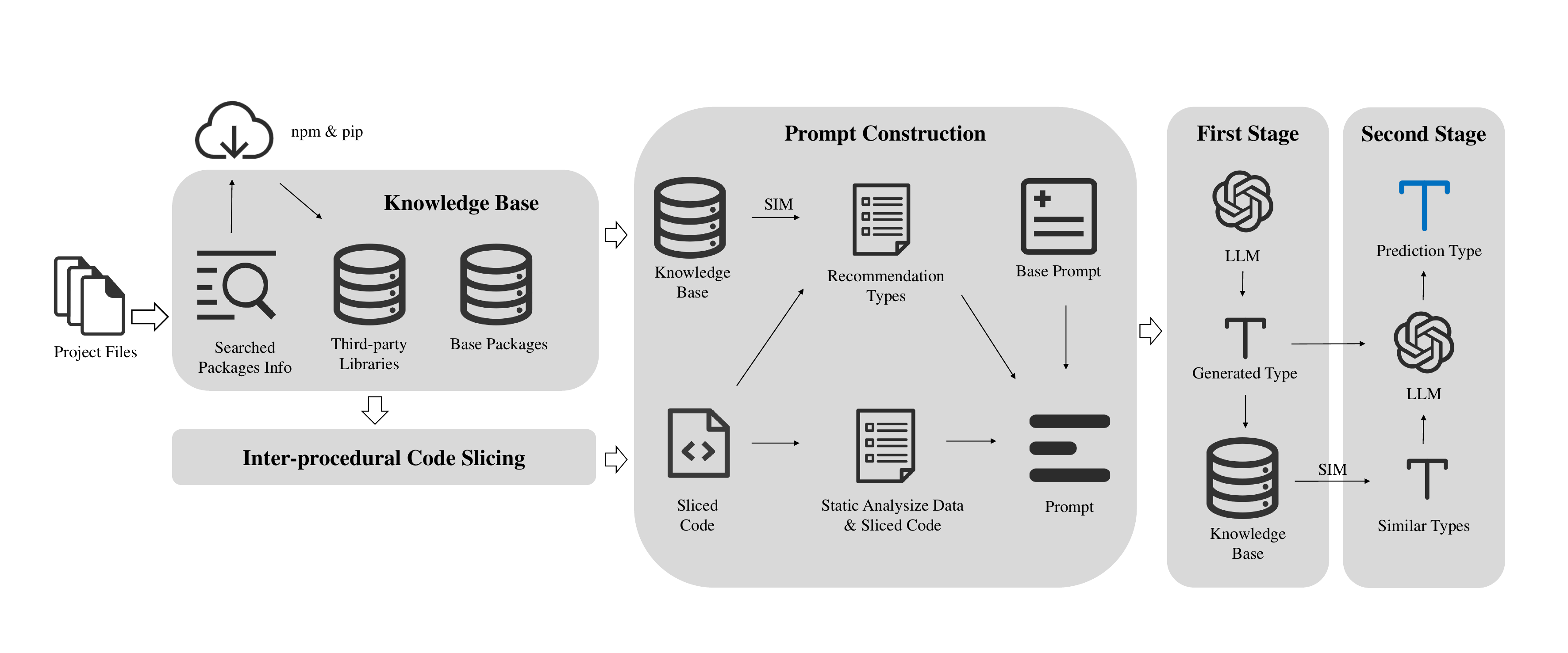}
    \caption{The Overview of TypePro.}
    \Description{The Overview of TypePro}
    \label{fig: flowchart}
\end{figure}

\subsection{Overview}
Figure ~\ref{fig: flowchart} illustrates the overall workflow of TypePro. First, the analysis process of TypePro relies on a knowledge base constructed from a large number of project files, incorporating domain knowledge of user-defined data types and third-party library types included in the project under analysis. When processing a target file, for each variable requiring type inference, TypePro performs inter-procedural code slicing to extract code snippets relevant to the variable at runtime as context. Based on the above two inputs, TypePro constructs a prompt containing (1) the code snippets obtained from slicing and (2) a list of candidate data types retrieved from the knowledge base. Finally, it leverages a LLMs to generate the inferred data type.

\subsection{Knowledge Base Construction}

To enable TypePro to support user-defined types and custom types from third-party libraries, TypePro collects all class definitions in the project files where the current file is located as user-defined types. Then, it examines the import statements to determine which source files are imported in the current file and adds the class definitions of these files to the list of user-defined types. For types from third-party libraries, we followed the existing practices of Peng et al.~\cite{peng2023generative} to download the top 5,000 most popular packages in the Python ecosystem (from \texttt{libraries.io}) and the top 5,000 packages in the TypeScript ecosystem (from \texttt{npm}). The knowledge base stores the following information closely related to the structure of user-defined data types: (1) class names, (2) the packages to which the classes belong, (3) public API signatures of these classes, and (4) the list of fields of these classes.

\subsection{Code Slicing}

\begin{algorithm}[t]
\caption{\revise{Code Slicing Process}}
\label{alg:IndividualSlicingProcess}
\KwIn{$P$: The target program;\\
$v$: The variable whose type is to be inferred.}
\KwOut{$slices$: A set of slices containing relevant statements for type inference.}
\tcp{4.3.1 System Dependency Graph Construction}
Initialize an empty list $PDG \gets []$\;
\ForEach{function $f$ in $P$}{
    $pdg_f = \text{buildPDG}(f)$\;
    Add $pdg_f$ to $PDG$\;
}
$G = \text{buildSDG}(PDG)$\; 
\tcp{4.3.2 Code Slicing Generation}
Identify the set of target variables $TV$ based on $v$\;
Initialize an empty list $slices \gets []$\;

\ForEach{$tv \in TV$}{
    Initialize the worklist with the pair $(tgtStmt, \{tgtStmt\})$\;
    
    \While{$worklist$ is not empty}{
        Remove $(s', slice)$ from $worklist$\;
        Mark $s'$ as visited\;
        
        \If{no unvisited statement $s$ with an edge $(s, s')$ in $G$}{
            Add $slice$ to $slices$\;
            \textbf{continue}\; 
        }
        
        \ForEach{unvisited statement $s$ with an edge $(s, s')$ in $G$}{
            \If{$s$ is a \textbf{def} or \textbf{use} of $tv$}{
                $slice \gets slice \cup \{s\}$\;
            }
            \Else{
                Add the pair $(s, slice)$ into $worklist$\;
            }
        }
    }
}
\Return $slices$\;
\end{algorithm}

TypePro performs inter-procedural code slicing to identify and analyze code segments essential to the target (e.g., variables, expressions, functions). Existing related work, such as TypeGen ~\cite{peng2023generative} and DLInfer~\cite{yan2023dlinfer}, also employs code slicing for type inference. However, as noted in Section 3, these methods exhibit deficiencies in handling inter-procedural dependencies, resulting in limited success rates when using sliced information as context for type inference. In contrast, TypePro’s inter-procedural slicing not only supplements context beneficial for type inference—such as signatures of used functions and call information—but also further incorporates other variables that exhibit dependency relationships with the target variable (either being depended on or depending on it). This approach provides richer and more effective contextual information for type inference.

\revise{Algorithm~\ref{alg:IndividualSlicingProcess} illustrates TypePro's slicing process, which is adopted from the process proposed by Horwitz et al.~\cite{horwitz1990interprocedural}. Given a target program $P$ as input, TypePro first constructs the System Dependence Graph (denoted as $G$) to support inter-procedural control-flow and data-flow analysis. Then, given the target variable $v$, TypePro identifies all statements in $G$ related to $v$ to provide the context for the LLM to infer the corresponding data type.}

\subsubsection{System Dependency Graph Construction}
\revise{TypePro first constructs $G$ for the target program $P$ (lines 1--4). Following the practices of existing slicing approaches \cite{yan2023dlinfer}, for each function defined in $P$, TypePro scans the code to extract data flow and control flow information to build the program dependency graph (PDG). Then, to construct the inter-procedural information of $G$, TypePro searches in PDG for call statements in the form of \texttt{ret1}, \texttt{ret2}, \ldots = \texttt{func}(\texttt{p1}, \texttt{p2}, \ldots), resolving the function \texttt{func} according to the following three rules:}

\begin{itemize}[leftmargin=*]

    \item \revise{\textbf{Function Name Matching. } TypePro first tries to match the name of the callee \texttt{func} in the call statement, and process only the functions with the same name further.}

    \item \revise{\textbf{Parameter Matching. }  TypePro further match the actual parameters at the call site with the formal parameters in the function signature. First, priority is given to explicit parameter names (e.g., keyword arguments) for alignment with the formal parameters of the candidate functions. If no explicit parameter names are provided, the original types of the actual parameters \texttt{(p1, p2,...)} (e.g., \texttt{number}, \texttt{float}, etc.) will be inferred and matched with the constraints of the candidate function's formal parameters.}

    \item \revise{\textbf{Return Value Matching. } In Python, a function may return more than one value. Therefore, TypePro scans the \texttt{return} statements within the body of function \texttt{func} to infer the actual number of return values, and checks whether this number matches the number of receiving variables (a.k.a, \texttt{ret1}, \texttt{ret2}, ...). For example, \texttt{return (x, y)} indicates that the function returns two values. If there are multiple \texttt{return} statements in the function body, TypePro takes the maximum number of return values among all the \texttt{return} statements as the number of return values for the function.}
\end{itemize}

\subsubsection{Code Slicing Generation}
After constructing $G$, TypePro takes as input the variable \( v \) for the project. In line 5, TypePro starts by expanding the set of target variables \( TV \) associated with \( v \) to enable precise type inference. Such the type of expansion is intended to support inter-procedural analysis and is governed by the following rules:

\begin{itemize}[leftmargin=*]
    \item \textbf{Fields:} For variables that are fields of a class, TypePro adds all the use statements of these fields to the $TV$.
    
    \item \textbf{Function Parameters:} Since function parameters lack explicit definitions, TypePro not only adds the function signature where the parameter appears to the $TV$, but also enumerates the uses of these functions within the project, adding the uses of these functions to $slice$ as well.
    
    \item \textbf{Return Values:} If the $v$ accepts the return value of a function (e.g., \texttt{ret1}, \texttt{ret2},...), then all return statements of that function are added to the $slice$.
    
    \item \textbf{Other Variables:} For the other variables, TypePro will directly add $v$ to $TV$.
\end{itemize}

\revise{Subsequently, for each identified $tv$ in $TV$, TypePro proceeds by performing backward and forward tracing on the identified $slice$ to collect all def and use statements (lines 7--19). Specifically, TypePro locates the statement $tgtStmt$ where $tv$ locates in $G$ (line 8), $tgtStmt$ also works as the initial slice for the given $tv$. Then, TypePro employs a worklist to progressively perform forward and backward propagation in $G$ to add the statements containing all related defs and uses of $tv$ into $slice$. 
This process continues until all def-use relations of the target variable and its related variables are fully traced, and the relevant statements are incorporated into the slice. The algorithm finally outputs the slices, which contain all the $slices$ for $v$ (line 20).}

\begin{figure}[htbp]
    \centering
    \includegraphics[width=0.93\textwidth,height=10.5cm]{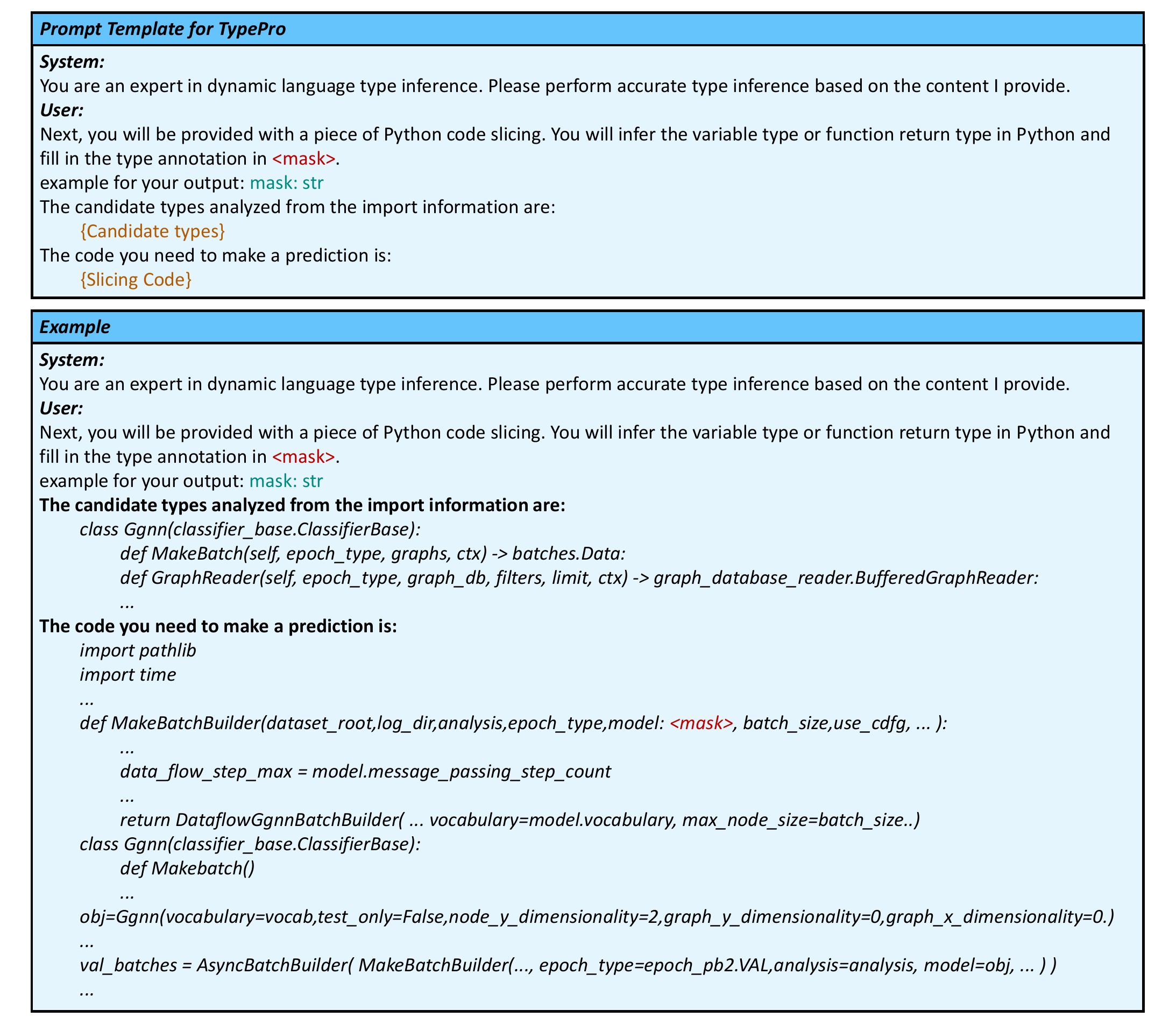}
    \caption{The prompt generated for the example in Figure \ref{fig:example code}}
    \Description{Prompt Example}
    \label{fig:Prompt Example}
\end{figure}

\subsection{Prompt Construction}
Finally, TypePro generates prompts and leverages a LLMs  for type inference. Figure~\ref{fig:Prompt Example} demonstrates the example prompt generated by TypePro to infer the data type of the variable \texttt{model} from Figure~\ref{fig:example code}. Within this prompt, the location of the variable whose type needs to be inferred is marked as \texttt{<mask>}, and the prompt incorporates the following two key pieces of information:
(1) The previously generated inter-procedural slice for the target variable;
(2) Candidate data types recommended from user-defined or third-party libraries.

As discussed in Section \ref{sec:motivation}, existing methods~\cite{Hityper, peng2023generative, wang2024tiger} struggle to effectively infer the data type of the variable \texttt{model} because the contextual information they rely on is insufficient to reconstruct the structural features of the data type (such as fields, public functions, etc.). TypePro addresses this limitation by recommending candidate data types whose structural information highly aligns with that inferred from the code slice, thereby enhancing the accuracy of type inference. Specifically, for a given candidate type type, TypePro calculates a matching score, which quantifies how closely the structure of type matches that of $tgtVar$. This score is computed as follows:

\begin{equation}
\label{equ:score}
Score(Var, type)
= \frac{\left| \left(F_{\text{Var}} \cup M_{\text{Var}}\right) \cap \left(F_{\text{type}} \cup M_{\text{type}}\right) \right|}
       {\left| F_{\text{Var}} \cup M_{\text{Var}} \right|}
\end{equation}

Specifically, 
$F_{Var}$ denotes the set of fields associated with $tgtVar$ that appear in $Slice$; $M_{Var}$ denotes the set of public functions associated with $tgtVar$ that appear in the Slice, $F_{type}$ denotes the set of fields defined in candidate type, and $M_{type}$ denotes the set of public functions defined in candidate type. TypePro selects up to $max_{candidate}$ (set to 5 in the evaluation) candidate data types from both the project itself and the knowledge base and then filter out data types whose similarity scores fall below the set minimum thresholds, we set the minimum threshold $threshold_{Score}$ for the $Score$ as 0.5. The detailed structural information of these candidate types is then explicitly listed in the prompt.

The aforementioned process constructs the initial prompt. However, LLMs may generate outputs in different formats. For example, (1) directly outputting the structure information of a data type, such as \{name: string, age: number\}, which is actually a user-defined type (e.g., \texttt{class Student}) that has already been included in the knowledge base, and (2) outputting the name of a data type but with typos (e.g., \texttt{Neme}, which is a typo for \texttt{Name}).
To reduce hallucination issues caused by the randomness of LLMs, TypePro executes a two-stage type generation process. In the first stage, TypePro provides the prompt generated through the above process to the LLMs and obtains its initial output. TypePro then repeats the above actions with reference to the output of the initial prompt. Specifically, for the first type of output, TypePro refers to Equation \ref{equ:score} to attempt to map it to a user-defined data type in the knowledge base. For the second type of output, TypePro considers the similarity of type names by using the BM25~\cite{BM25} algorithm, and selects the most similar type from the knowledge base as the output.

\subsection{\revise{Discussion}}



\revise{Note that TypePro’s effectiveness may be affected by limitations in SDG construction. As discussed in Section 4.3.2, inferring inter-procedural information relies on matching function names, parameters, and return values. For certain call statements, TypePro may identify multiple candidate functions and conservatively provide all of them to the LLM. While this ensures completeness, it may introduce imprecision that affects type inference. To evaluate the impact of the aforementioned limitations, we sampled 2,500 variables from the ManyTypes4Py and ManyTypes4TypeScript datasets. In ManyTypes4Py, 2,273 variables involved slicing with a single matching function, achieving \(89.93\%\) precision; the remaining 227 with multiple matches achieved \(89.42\%\). In ManyTypes4TypeScript, 2,321 variables had a single match with \(85.61\%\) precision, and 179 had multiple matches with \(85.47\%\) precision. These results indicate that the above process has limited impact on TypePro’s performance.}

\revise{Another threat is that the proposed slicing process does not support code generation mechanisms in Python and TypeScript, such as calls to the \texttt{eval} function, which are rarely used. In the ManyTypes4Py dataset, there are 210 \texttt{eval} calls across 122 files, while in the ManyTypes4TypeScript dataset, there are only 5 \texttt{eval} calls in 2 files. As stated in existing work~\cite{salis2021pycg,guarnieri2009gatekeeper, staicu2018understanding}, such dynamic features limit the effectiveness of static analysis approaches. We plan to explore incorporating runtime information to better handle these features in future work.}
\section{Evaluation}

\begin{table}[t]
\centering
\caption{Dataset for Type Inference. Specifically, \textbf{Arg} refers to function parameter types, \textbf{Ret} refers to function return type, \textbf{Var} refers to local and global variables, \textbf{Gen} refers to generic types, \textbf{Ele} refers to elementary types (e.g., \texttt{str} and \texttt{int}), and \textbf{Usr} denotes types defined in the project or third-party libraries.}
\renewcommand{\arraystretch}{1.2} 
\scalebox{0.9}{
\begin{tabular}{lcccccccc}
\toprule
\multirow{2}{*}{\textbf{Language}} &\multirow{2}{*}{\textbf{Dataset}} & \multirow{2}{*}{\textbf{Total}} & \multicolumn{6}{c}{\textbf{Categories}} \\
\cline{4-9}
                         &          &              & \textbf{Arg} & \textbf{Ret} & \textbf{Var} & \textbf{Ele} & \textbf{Gen} & \textbf{Usr} \\
\toprule
\multirow{6}{*}{\textbf{Python}} &\multirow{2}{*}{Training Set} & 226,767 & 44,276 & 20,395 & 162,096 & 119,742 & 62,268 & 44,757 \\
                              && 100\%   & 19.52\% & 8.99\%  & 71.48\%  & 52.80\%  & 27.46\% & 19.74\% \\
\cline{2-9}
&\multirow{2}{*}{Test Set}      & 101,392  & 20,885 & 9,393  & 71,114  & 52,869  & 28,227 & 20,296 \\
                              & & 100\%   & 20.60\% & 9.26\%  & 70.14\%  & 52.14\%  & 27.84\% & 20.02\% \\
\cline{2-9}
&\multirow{2}{*}{Sampled Test Set} & 11,029 & 2,269  & 958    & 7,802   & 5,684   & 3,312  & 2,033  \\
                                  && 100\%  & 20.57\% & 8.69\%  & 70.74\%  & 51.54\%  & 30.03\% & 18.43\% \\
\midrule
\multirow{4}{*}{\textbf{TypeScript}}&\multirow{2}{*}{Sampled Training Set} & 260,767 & 106,571 & 42,780 & 111,416 & 246,784 &-& 13,983\\
                                     & & 100\%   & 40.87\% & 16.41\%  & 42.73\%  & 94.64\%  &-& 5.36\%  \\
\cline{2-9}
&\multirow{2}{*}{Sampled Test Set} & 30,805 & 13,019  & 4,055    & 13,731   & 29,731 &-  & 1,074   \\
                                 & & 100\%  & 42.26\% & 13.16\%  & 44.57\%  & 96.51\% &- & 3.49\%  \\
\bottomrule
\end{tabular}
}

\label{tab: datasets}
\end{table}

We evaluated TypePro by answering the following research questions (RQs):
\begin{itemize}[leftmargin=*]
    \item \textbf{RQ1:} How does TypePro perform on Python datasets compared to other baselines?
    \item \textbf{RQ2:} How robust is TypePro in inferring types for different categories of variables, including elementary types, generic types, and user-defined types?
    \item \revise{\textbf{RQ3:} Can the results of TypePro generalized to different types of LLMs?}
    \item \textbf{RQ4:} What is the contribution of each component of TypePro (i.e., \textbf{code slicing} and \textbf{type recommendation}) to its overall effectiveness?
    \item \textbf{RQ5:} How effective is TypePro on inferring data types in TypeScript programs?
\end{itemize}

\subsection{Evaluation Datasets}
We utilized the ManyTypes4Py~\cite{mir2021manytypes4py} and ManyTypes4TypeScript~\cite{Manytypes4typescript} datasets for our experiments. Specifically, ManyTypes4Py is a widely adopted dataset containing 869K annotations of 5,382 Python projects for type inference. Following the practices of recent studies~\cite{peng2023generative}, we partitioned the dataset into training and testing sets using a 70-30 split and further selected data from 100 projects within the test set for evaluation.
On the other hand, the ManyTypes4TypeScript dataset is designed for type inference in TypeScript. ManyTypes4Typescript contains 9M+ data annotations in 13,953 projects, which is ten times the size of ManyTypes4Py. We extracted 10\% of the pre-partitioned training set from it as the training data and further selected data from 100 projects within the test set for evaluation. The statistics of selected training and test dataset are shown in Table \ref{tab: datasets}. Note that the ManyTypes4TypeScript dataset does not include data annotations for generic types (e.g., generic types such as \texttt{Array<T>} and \texttt{Map<K, V>} are annotated simply as \texttt{array} or \texttt{map}). 

\subsection{Baselines}
We selected the following research methods in type inference as our baselines.  
\begin{itemize}[leftmargin=*]
    
\item \textbf{Type4Py~\cite{mir2022type4py}}: A supervised learning-based classification method that constructs large-scale Python type clusters and categorizes programs into different clusters for type prediction. For Type4Py, we directly used the Docker image released by the authors to perform inference on the test set.

\item \textbf{HiTyper~\cite{Hityper}}: A method that builds a Type Dependency Graph (TDG) to describe variable composition and dependencies, combined with a similarity-based model for type inference. For HiTyper, we utilized Type4Py as its inference model.

\item \textbf{TypeGen~\cite{peng2023generative}}: An approach that performs code slicing via TDG, constructs a chain-of-thought and examples using natural language descriptions of static analysis knowledge, and then leverages a LLMs to generate types. For TypeGen, we followed the default parameters set by the authors, preprocessed the dataset according to their methodology, and then conducted inference on the test set.

\item \textbf{TIGER~\cite{wang2024tiger}}: A two-stage type inference method for Python that employs a trained generative model along with a similarity computation model to rank and recommend plausible types. For TIGER, we adhered to the authors' approach, training the model on our training set and subsequently performing predictions.
\end{itemize}

Note that we omitted rule-based approaches as they are unlikely to outperform machine learning-based methods in terms of performance and generalization on these datasets. Furthermore, we compared TypePro with three cloze-style type inference models: CodeT5~\cite{codeT5}, CodeT5+~\cite{codeT5+}, and UnixCoder~\cite{guo2022unixcoderunifiedcrossmodalpretraining}. We did not include DLInfer~\cite{yan2023dlinfer} in our baselines because its type inference process relies on PySlicer, a static slicing tool for Python that has not been open-sourced. For CodeT5~\cite{codeT5}, CodeT5+~\cite{codeT5+}, and UnixCoder~\cite{guo2022unixcoderunifiedcrossmodalpretraining}, we downloaded the initial weights from the Hugging Face Hub~\cite{huggingface2025hub} and fine-tuned them on our training set. 

\subsection{Metrics}
We employ the following metrics from prior work~\cite{mir2022type4py, peng2023generative,wang2024tiger,Hityper} to evaluate the performance of TypePro and the baseline methods:
\begin{itemize}[leftmargin=*]
    \item \textbf{Exact Match (EM):} This metric calculates the proportion of type predictions made by a method that exactly match the developer-provided type annotations.

    \item \textbf{Base Match (BM):} This metric calculates the proportion of type predictions that partially match the developer-provided type annotations.
\end{itemize}

Formally, we parse Python or TypeScript tag types $T_{label}$ and the predicted types from a method $T_{pre}$ into type sets $Set_{\text{label}}$ and $Set_{pre}$. 

$$
T_{label}\ \rightarrow \ Set_{label}\ =\ Set\left( T_1,T_2,\cdots T_n \right) 
$$
$$
T_{pre}\ \rightarrow \ Set_{pre}\ =\ Set\left( T_1,T_2,\cdots T_n \right) 
$$

For example, the Python type \texttt{Union[str, int]} is parsed as $Set(\texttt{str}, \texttt{int})$, and the TypeScript type \texttt{\{a?:number,  b:string\}} is parsed as $Set(\texttt{\{a:number,b:string\}}, \texttt{\{b:string\}})$.
When the two type sets are identical, we consider Exact Match to be satisfied. If the two sets have any overlap, we consider Base Match to be satisfied.

$$
Set_{label}\ =\ Set_{pre}\ \Rightarrow \ EM
$$
$$
Set_{label}\ \cap \ Set_{pre}\ \ne \ \phi \ \Rightarrow \ BM
$$
 
Notably, for generic types in Python and TypeScript, we consider predictions to satisfy BM as long as they belong to the same generic category. For example, in Python, \texttt{List[str]} and \texttt{List[int]} are considered a base match; similarly, in TypeScript, \texttt{Array<string>} and \texttt{Array<number>} are considered a base match.

For the \texttt{any} type:
If the annotated type in the dataset is \texttt{any}, any predicted type is considered to satisfy both EM and BM.
If the predicted type is \texttt{any}, but the annotated type in the dataset is any specific type (other than \texttt{any}), it is considered to satisfy neither EM nor BM.

Furthermore, we utilized MRR (Mean Reciprocal Rank) as an evaluation metric~\cite{MRRs,mir2022type4py}. This is a commonly used indicator to assess the performance of models such as search engines or recommendation systems, particularly suited for query-based problems like information retrieval and recommendation tasks. The score for each data point is calculated by taking the reciprocal of the rank of the first correct answer. The formula is as follows:
    \begin{equation}
    \label{eq:mrr}
	MRR@n = \frac{1}{|D|} \sum_{i=1}^{|D|} \frac{1}{rank_i} (rank_i \in [1, n])
    \end{equation}
where $|D|$  denotes the total number of data, and $\operatorname{rank}_i$ represents the rank position of the first correct answer for the $i$-th query, $MRR@n$ represents the retrieval of data ranked from 1 to n, and $rank_i$ falls within the range $[1, n]$.

TypePro utilized the API provided by OpenAI and selected GPT-4.1 as the generative model, which is one of the state-of-the-art LLMs. We set the temperature to 0.2 and top\_p to 0.3, generating 20 samples for each target variable. The final prediction was determined by ranking the generated results based on their frequency of occurrence, This is the same approach as other studies~\cite{peng2023generative}. All experiments were conducted on a Linux machine (Ubuntu 24.04) equipped with a 128-core Intel Xeon Gold 6338 CPU @ 3.200GHz and two NVIDIA RTX A6000 GPUs.

\begin{table}[t]
\centering
\caption{The ManyTypes4Py results of TypePro and baselines in terms of \textbf{Var}, \textbf{Ret}, and \textbf{Arg}. The data point with a dark blue background indicates the best result. The data point with a light blue background indicates the second-best result.}
\tiny
\renewcommand{\arraystretch}{1.5} 
\scalebox{0.95}{
\begin{tabular}{lcc|cccc|cccc|cccc|cccc}
\toprule
\multirow{2}{*}{\textbf{Metric}}& \multirow{2}{*}{\textbf{Approach}} & \multirow{2}{*}{\textbf{Catalog}} & \multicolumn{4}{c}{\textbf{Top-1}} &\multicolumn{4}{c}{\textbf{Top-3}} & \multicolumn{4}{c}{\textbf{Top-5}}& \multicolumn{4}{c}{\textbf{MRR@5}}\\
\cline{4-7} \cline{8-11} \cline{12-15} \cline{16-19}
                         & &                        &   \textbf{Var} & \textbf{Ret} & \textbf{Arg} & \textbf{All} & \textbf{Var} & \textbf{Ret} & \textbf{Arg} & \textbf{All} &\textbf{Var} & \textbf{Ret} & \textbf{Arg} & \textbf{All} &\textbf{Var} & \textbf{Ret} & \textbf{Arg} & \textbf{All}\\
    \toprule
\multirow{8}{*}{EM}& \multirow{1}{*}{HiTyper} & \multirow{2}{*}{CLS} & 82.2 & 60.7 & 67.3 & 76.2& 85.5 & 64.5 & 77.3 & 81.4  & 85.9 & 65.1 & 78.3 & 81.9&87.0 & 70.4 & 68.5& 81.7\\
&\multirow{1}{*}{Type4Py} &  & 62.2 & 41.8 & 68.1 & 61.9 & 70.8 & 51.1 & 81.1 & 71.5 & 72.7 & 53.4 & 81.7 & 73.1&66.5 & 46.7 & 74.3& 66.6\\
\cline{2-19}
&\multirow{1}{*}{TypeGen} & \multirow{4}{*}{GEN} & 72.6 & 62.1 & 67.4 & 70.7 & 81.2 & 71.9 & 79.0 & 80.0 & 82.8 & 73.9 & 81.4 & 81.7&76.9 & 67.1 &73.2 & 75.3 \\
&\multirow{1}{*}{CodeT5} &  & 84.7 & 62.9 & 64.6 & 78.7  & 89.4 & 69.9 & 74.2 & 84.6 & 91.0 & 72.7 & 76.3 & 86.4&87.2 & 66.8 & 69.4& 81.6\\
&\multirow{1}{*}{Unixcoder} &  & 83.2 & 65.8 & 60.9 & 77.1 & 90.5 & 75.1 & 75.9 & 86.2 & \cellcolor{myc!50}92.3 & 77.0 & 79.2 & 88.3&86.9 & 70.4 & 68.5& 81.7\\
&\multirow{1}{*}{CodeT5+} &  & 84.6 & 65.4 & 63.4 & 78.6 & 88.8 & 71.8 & 73.3 & 84.1 & 90.5 & 73.9 & 75.6 & 86.0&86.9 & 68.9 & 68.4& 81.5\\
\cline{2-19}
&\multirow{1}{*}{TIGER} & \multirow{2}{*}{GEN+SIM} & \cellcolor{myc!50}85.1 & \cellcolor{myc!50}70.6 & \cellcolor{myc!50}76.0 & \cellcolor{myc!50}81.9 & \cellcolor{myc!50}90.7 & \cellcolor{myc!50}81.2 & \cellcolor{myc!50}89.3 & \cellcolor{myc!50}89.6  & 91.5 & \cellcolor{myc!50}83.3 & \cellcolor{myc!50}91.4 & \cellcolor{myc!50}90.8&\cellcolor{myc!50}87.8 & \cellcolor{myc!50}75.8 & \cellcolor{myc!50}82.7& \cellcolor{myc!50}85.7\\
&\multirow{1}{*}{TypePro} &  & \cellcolor{myc2!50}89.6 & \cellcolor{myc2!50}88.9 & \cellcolor{myc2!50}87.2 & \cellcolor{myc2!50}88.9 & \cellcolor{myc2!50}93.1 & \cellcolor{myc2!50}91.6 & \cellcolor{myc2!50}91.3 & \cellcolor{myc2!50}92.4 &\cellcolor{myc2!50}93.2 & \cellcolor{myc2!50}91.9 & \cellcolor{myc2!50}91.7 & \cellcolor{myc2!50}92.7&\cellcolor{myc2!50}91.1&\cellcolor{myc2!50}90.2&\cellcolor{myc2!50}89.1&\cellcolor{myc2!50}90.6\\
\midrule
\multirow{8}{*}{BM}& \multirow{1}{*}{HiTyper} & \multirow{2}{*}{CLS} & 90.3 & 80.6 & 74.7 & 85.2  & 93.8 & 84.9 & 84.6 & 90.5 & 94.4 & 85.2 & 85.8 & 91.3&91.9 & 80.6 & 75.1& 87.5\\
&\multirow{1}{*}{Type4Py} &  & 69.3 & 45.2 & 70.0 & 67.6 & 79.5 & 55.8 & 83.2 & 78.4 & 81.4 & 58.2 & 84.4 & 80.2&74.4 & 50.6 &76.4&73.0\\
\cline{2-19}
&\multirow{1}{*}{TypeGen} & \multirow{4}{*}{GEN} & 85.5 & 70.7 & 75.0 & 82.1 & 90.0 & 80.1 & 85.1 & 88.1 & 90.6 & 81.7 & 86.4 & 88.9 &87.7 & 75.5 & 79.9& 85.1\\
&\multirow{1}{*}{CodeT5} &  & 90.9 & 70.6 & 72.8 & 85.4 & 94.4 & 78.3 & 81.8 & 90.4 &95.6 & 80.6 & 84.2 & 91.9&\cellcolor{myc!50}92.7& 74.6& 77.4&87.9\\
&\multirow{1}{*}{Unixcoder} &  & 89.1 & 76.8 & 68.1 & 83.7 & \cellcolor{myc!50}94.5 & 84.6 & 81.8 & 91.0  & \cellcolor{myc2!50}96.2 & 86.5 & 85.1 & 93.1&91.9 & 80.6 & 75.1& 87.5\\
&\multirow{1}{*}{CodeT5+} &  &\cellcolor{myc!50}90.9 & 75.5 & 70.7 & 85.4  & 94.2 & 82.6 & 79.5 & 90.2 & 95.5 & 84.9 & 82.2 & 91.9 &\cellcolor{myc!50}92.7 & 79.3 & 75.3& 87.9\\
\cline{2-19}
&\multirow{1}{*}{TIGER} & \multirow{2}{*}{GEN+SIM} & 90.8 & \cellcolor{myc!50}78.9 & \cellcolor{myc!50}82.6 & \cellcolor{myc!50}88.1 & 93.7 & \cellcolor{myc!50}88.7 & \cellcolor{myc2!50}93.0 & \cellcolor{myc!50}93.1 & 94.3 & \cellcolor{myc!50}90.0 & \cellcolor{myc2!50}94.7 & \cellcolor{myc!50}94.0&92.3 & \cellcolor{myc!50}83.6 & \cellcolor{myc!50}87.9 & \cellcolor{myc!50}90.6\\
&\multirow{1}{*}{TypePro} &  &\cellcolor{myc2!50} 93.5 & \cellcolor{myc2!50}91.2 & \cellcolor{myc2!50}88.1 & \cellcolor{myc2!50}92.1  & \cellcolor{myc2!50}95.6 & \cellcolor{myc2!50}94.1 & \cellcolor{myc!50}91.7 & \cellcolor{myc2!50}94.6  & \cellcolor{myc!50}95.7 & \cellcolor{myc2!50}94.2 & \cellcolor{myc!50}92.1 & \cellcolor{myc2!50}94.8&\cellcolor{myc2!50}94.5&\cellcolor{myc2!50}92.5&\cellcolor{myc2!50}89.8&\cellcolor{myc2!50}93.3\\
\bottomrule
\end{tabular}}
\label{tab: RQ1 result1}
\end{table}

\subsection{Results of RQ1}

To answer RQ1, we evaluate the effectiveness of TypePro in type inference of Python by comparing the results with baselines on ManyTypes4Py~\cite{mir2021manytypes4py}.
We compared two classification-based methods, HiTyper~\cite{Hityper} and Type4Py~\cite{mir2022type4py}, as well as generative models such as CodeT5~\cite{codeT5}, CodeT5+~\cite{codeT5+} and UnixCoder~\cite{guo2022unixcoderunifiedcrossmodalpretraining}. TypeGen is categorized as the GEN-based approach, as LLMs are also considered as a form of generative model.

Table \ref{tab: RQ1 result1} presents the results. As we can see, TypePro achieved an 88.9\%, 92.7\%, and 92.7\% accuracy in Top-1, Top-3, and Top-5 Exact Match, representing an improvement of 7.0, 2.8, and 1.9 percentage points over the second-best method, TIGER. In detail, TypePro achieved the best results across all aspects of exact match in Var, Ret, and Arg. An interesting observation is that, for Top-1 Exact Match, TypePro showed remarkable improvements in inferring Ret and Arg, outperforming TIGER by 18.3 and 11.2 percentage points, respectively. As stated by existing work\cite{parnin2011automated, mir2022type4py}, developers are more likely to use the first suggestion provided by a tool, which highlight the effectiveness of TypePro in achieving the highest success rate without requiring additional manual effort.

For Base Match, TypePro also achieves a Top-1 accuracy of 92.1\%, which is a 4 percentage point improvement over the second-best method. Specifically, in terms of Top-3 and Top-5 results, TypePro ranks either first or second across different groups. This demonstrates the robustness of TypePro regarding effectiveness under different evaluation settings.

\begin{figure}[t]
    \centering
    \begin{minipage}[b]{\textwidth}
        \centering
        \includegraphics[width=0.9\linewidth,height=9cm]{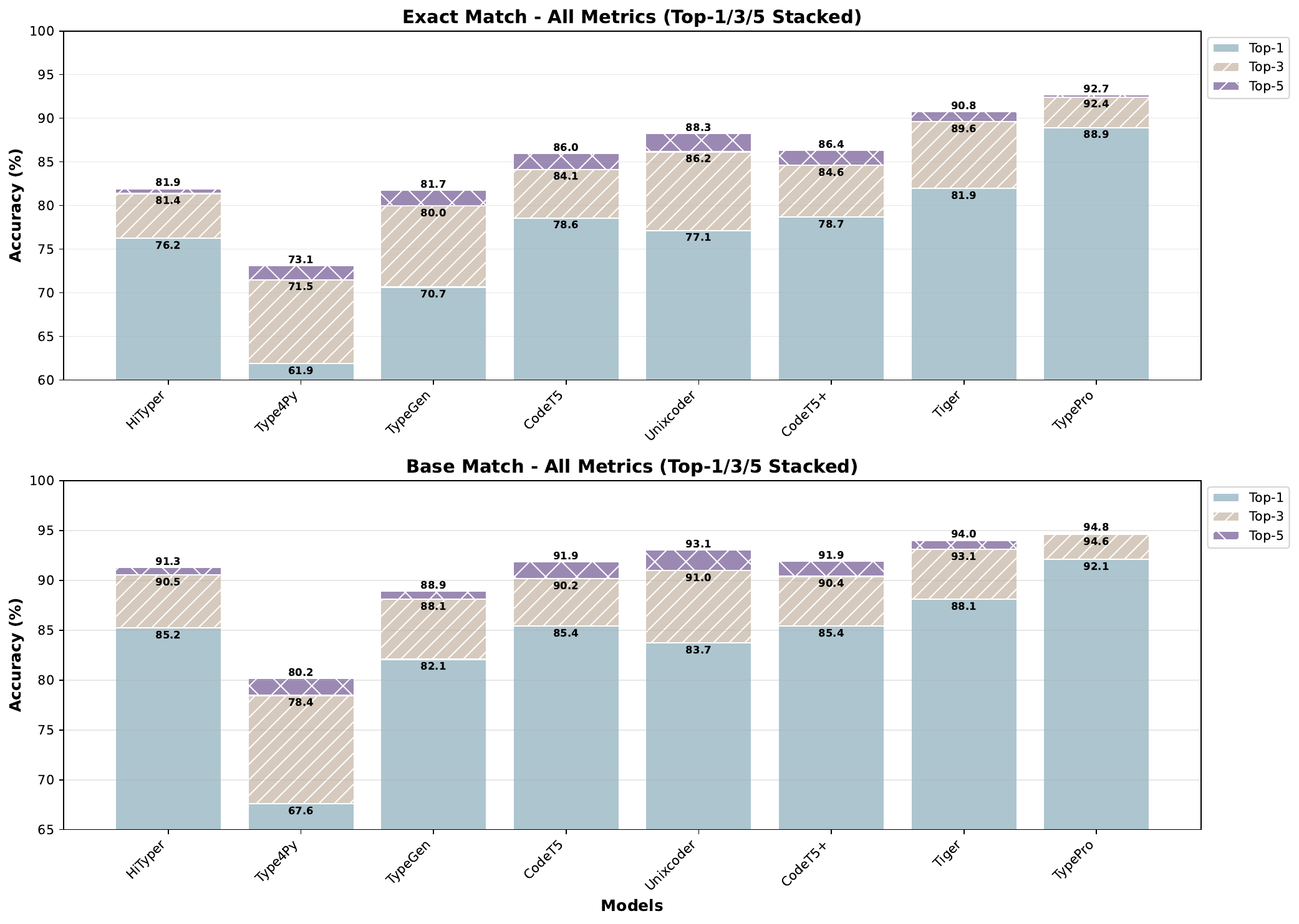}
        \caption{Correct answer distribution between TypePro and baselines}
        \label{fig:stack fig}
    \end{minipage}
\end{figure}

\subsubsection{Comparison with Classification-Based Approaches}
As shown in Table \ref{tab: RQ1 result1}, TypePro clearly outperforms CLS approaches, including HiTyper and Type4Py, across all type categories in terms of Exact Match and Base Match metrics. Notably, TypePro achieves improvements of 12.8 and 27.1 percentage points over HiTyper and Type4Py, respectively, in Top-1 Exact Match.

The main reason for these results is that elementary types are typically included in the vocabularies of CLS approaches' pre-trained models and have a large number of training samples. However, such a large amount of training data does not include information generated by inter-procedural analysis. Therefore, when applying HiTyper and Type4Py to Ret and Arg categories, their performance drops significantly. For example, the accuracy of Type4Py on Ret category is only 41.8\%. In contrast, TypePro achieves success rates of 88.9\% and 87.2\% on these categories, respectively, highlighting its advantage over CLS approaches.

\subsubsection{Comparison with Generation-Based Approaches}
As shown in Table~\ref{tab: RQ1 result1}, the generation-based (GEN) approaches CodeT5 and CodeT5+ achieved Top-1 accuracies of 78.7\% and 78.6\%, respectively, while UnixCoder and TypeGen reached 77.1\% and 70.7\%, respectively. For the same reason mentioned above, these models also performed well in the \textit{Var} category; for example, CodeT5 achieved an accuracy rate of 84.7\% in this category, largely because variables constitute a significant portion of the training data (71.48\%). However, the accuracy of these three methods in other categories drops to below 70\%. In contrast, TypePro’s inter-procedural code slicing process incorporates structural information of types, giving it an advantage in inferring the \textit{Ret} and \textit{Arg} categories. It is worth noting that although TypeGen pre-extracts some third-party library types and user-defined types for inclusion in its prompts to the LLMs, it only provides class names without structural information. This lack of contextual detail leads to reduced success rates.

\subsubsection{Comparison with TIGER}

As shown in Table~\ref{tab: RQ1 result1}, TIGER achieved a Top-1 exact match rate of 81.9\% on our test set, while TypePro outperformed TIGER by 7.9 percentage points in overall accuracy. Specifically, although the performance on Var types was similar, TIGER focuses on variable names and contextual textual similarity, whereas TypePro leverages inter-procedural analysis to obtain contextual information of the target variable and provides it to the LLM. This enables TypePro to achieve improvements of 17.3 and 11.2 percentage points over TIGER for Ret and Arg types, respectively.

For the Base Match task, although we can see that TypePro achieves overall better Top-1, Top-3, and Top-5 performance than TIGER on the test set, its performance in the Arg group for infer Top-3 and Top-5 only reaches 91.7\% and 92.1\%, which is 1.3 and 2.6 percentage points lower than TIGER, respectively. According to our observations, since TypePro is an LLM-based type inference method, when we configure the LLM to output more diverse results, it often produces hallucinations, leading to a decline in performance for Top-3 and Top-5 predictions. Nevertheless, TypePro still demonstrates a significant advantage over TIGER in terms of Top-1 performance.

\subsubsection{MRR}
As we can see in Table ~\ref{tab: RQ1 result1}, In terms of MRR@5, TypePro achieves 90.6 and 93.2, outperforming all the baselines we compared. It can be observed that TypePro significantly outperforms baselines in MRR@5 for Ret and Arg category, with 14.4 and 6.4 percentage point of improvement on Ret and Arg, respectively. For Var category, the difference between TypePro and the second-best approach is minimal, as these categories are relatively straightforward and well-represented in the training set. 
Moreover, Figure \ref{fig:stack fig} shows the distribution of correct answers between TypePro and baselines. Moreover, Figure \ref{fig:stack fig} shows the distribution of correct answers between TypePro and the baselines. As we can see, on the test dataset, TypePro ranks the correct answers in the Top-1, Top-3, and Top-5 for 88.9\%, 92.4\%, and 92.7\% of the data, respectively, all of which surpass the baselines. This demonstrates that TypePro consistently prioritizes the correct data types more effectively, further validating its advantages in ranking accuracy across different baselines.

\revise{\subsubsection{Efficiency} On the ManyTypes4Py dataset, TypePro achieves an average runtime of 2.46 seconds per sample (range: 1.38–8.38 seconds). The slicing operation itself averages only 0.50 seconds (range: 0.33–0.72 seconds), indicating that the more time-consuming part is not code slicing but rather LLM API calls which inevitably introduce network latency. Additionally, statistical analysis of prompt lengths shows an average of 1.96K tokens (range: 0.83K–10.45K). These results demonstrate that both the runtime and token consumption of TypePro fall within an acceptable range, reflecting stable and efficient performance.}

\revise{\subsubsection{Sensitivity Analysis} We conducted a sensitivity analysis on the selection of threshold $Score$ using the ManyTypes4Py dataset, with values set at 0.1, 0.3, 0.5, 0.7, and 0.9. The corresponding Top-1 exact match accuracies were 86.4\%, 87.2\%, 88.9\%, 86.9\%, and 86.9\%, respectively. The best performance was achieved when $Score$ was set to 0.5. Specifically, a low $Score$ value leads to too many irrelevant type information, while a high $Score$ makes the strategy too conservative and may exclude valid types that contain optional fields, both of which adversely affect performance.}

\begin{mybox}
    \textbf{Answer of RQ1:} TypePro achieves higher accuracies and MRR@5 compared to all the baselines. Notably, TypePro improves the Top-1 Exact Match by 7.1 percentage points over the second-best approach, TIGER. This demonstrates the effectiveness of TypePro in various scenarios, especially in the scenarios where fully automation is required (e.g., compilation optimization).
\end{mybox}



\subsection{Results of RQ2}
\begin{table}[t]
\centering
\caption{Comparison results with baselines on the three data types \textbf{Ele}, \textbf{Gen}, and \textbf{Usr}.}
\tiny
\renewcommand{\arraystretch}{1.5} 
\scalebox{1.12}{
\begin{tabular}{lc|ccc|ccc|ccc|ccc}
\toprule
 \multirow{2}{*}{\textbf{Approach}} & \multirow{2}{*}{\textbf{Catalog}} & \multicolumn{3}{c}{\textbf{Top-1}} &\multicolumn{3}{c}{\textbf{Top-3}} & \multicolumn{3}{c}{\textbf{Top-5}}& \multicolumn{3}{c}{\textbf{MRR@5}}\\
\cline{3-5} \cline{6-8} \cline{9-11} \cline{12-14}
                          &                       &   \textbf{Ele} & \textbf{Gen} & \textbf{Usr}  & \textbf{Ele} & \textbf{Gen} & \textbf{Usr}  &\textbf{Ele} & \textbf{Gen} & \textbf{Usr} &\textbf{Ele} & \textbf{Gen} & \textbf{Usr} \\
    \toprule
 \multirow{1}{*}{HiTyper} & \multirow{2}{*}{CLS} &86.6 & 63.6 & 66.9 & 92.0 & 67.5 & 73.3  & 92.7 & 68.0 & 73.6& 89.8 & 75.4 & 69.4 \\
\multirow{1}{*}{Type4Py} &  & 79.6 & 42.0 & 45.9  & 88.1 & 54.5 & 53.9 & 89.3 & 57.6 & 54.3 & 83.8 &48.2 &49.7 \\
\cline{1-14}
\multirow{1}{*}{TypeGen} & \multirow{4}{*}{GEN} & 84.2 & 44.8 & 74.8 & 88.9 & 62.4 & \cellcolor{myc!50}83.4  & 89.8 & 66.3 & 84.4 & 86.5 &53.7 & 79.0\\
\multirow{1}{*}{CodeT5} &  & 88.3 & 68.9 & 67.6 & 92.3 & 76.7 & 75.8  & 93.7 & 79.0 & 77.7& 90.5 & 72.3 &71.8 \\
\multirow{1}{*}{Unixcoder} &  & 86.3 & 69.8 & 63.5 & \cellcolor{myc!50}93.2 & 80.8 & 75.3 & \cellcolor{myc!50}94.8 & 83.5 & 77.8& 89.8 & 75.4 & 69.4 \\
\multirow{1}{*}{CodeT5+} &  & \cellcolor{myc!50}88.9 & 68.6 & 65.8 & 92.9 & 75.2 & 74.2  & 94.1 & 77.7 & 76.8 & \cellcolor{myc!50}91.0 & 72.2 &70.3 \\
\cline{1-14}
\multirow{1}{*}{TIGER} & \multirow{2}{*}{GEN+SIM} & \cellcolor{myc!50}88.9 & \cellcolor{myc!50}71.6 & \cellcolor{myc!50}79.3 & 91.7 & \cellcolor{myc!50}84.4 & \cellcolor{myc2!50}92.3  & 91.8 & \cellcolor{myc!50}86.7 & \cellcolor{myc2!50}94.7& 90.3 & \cellcolor{myc!50}77.9 & \cellcolor{myc!50}85.7 \\
\multirow{1}{*}{TypePro} &  & \cellcolor{myc2!50}93.2 & \cellcolor{myc2!50}82.6 & \cellcolor{myc2!50}87.6  & \cellcolor{myc2!50}95.0 & \cellcolor{myc2!50}88.4 & \cellcolor{myc2!50}92.3 &\cellcolor{myc2!50}95.0 & \cellcolor{myc2!50}88.7 & \cellcolor{myc!50}92.9&\cellcolor{myc2!50}93.9&\cellcolor{myc2!50}85.4&\cellcolor{myc2!50}89.8 \\
\bottomrule
\end{tabular}
}
\label{tab: RQ2 result}
\end{table}

To answer RQ2, we compared the performance of TypePro against various baseline methods across different data types. The data was categorized into three classes: elementary types (Ele), generic types (Gen), and user-defined types (Usr). We recorded the exact match results for TypePro and the baselines at Top-1, Top-3, and Top-5. 

The results are shown in Table~\ref{tab: RQ2 result}. Overall, TypePro outperforms the baselines with Top-1 accuracies of 93.2\%, 82.6\%, and 87.6\% for Ele, Gen, and Usr, respectively. Compared to the baselines, TypePro showed improvements in Top-1 accuracy, particularly in the Gen and Usr categories, with improvements of 11.0 and 8.3 percentage points, respectively.

However, in the Top-3 and Top-5 scenarios, TIGER slightly outperforms TypePro in user-defined types. The main reason is that, during the candidate type ranking process, TypePro only recommends data types with a fitness score $Score > 0.5$, which may result in fewer than three or five data types being recommended. In such cases, TypePro relies solely on semantic inference from code slices for data type prediction, which can easily lead to hallucinations. Future research directions could guide large language models in type inference to not only consider the structural information of types, but also design appropriate guidance mechanisms to help models understand the semantic relations between the inferred data type and the context, thereby further enhancing the effectiveness of LLM-based approaches.


As for the Ele category, we observe that the three code generation models (CodeT5, UnixCoder, CodeT5+) achieve over 80\% accuracy. This is because Ele types dominate the training dataset (52.8\% in ManyTypes4Py). When fine-tuning these models, the inherent bias in the data distribution led to their performance on Ele types. Additionally, compared to generation-based methods, classification-based methods clearly underperformed on user-defined types. For example, Type4Py achieved only 45.8\% Top-1 accuracy on Usr types, indicating that vocabulary-dependent classification methods struggle to handle types not present in the training vocabulary.

The MRR@5 results demonstrate that TypePro still achieved the best overall performance across the Ele, Gen and Usr. Although TIGER performed slightly better on Top-5 user-defined types, TypePro’s MRR@5 remained higher than TIGER’s. This is because TIGER’s Top-1 accuracy was relatively low and despite its increase from Top-1 to Top-3, Top-1 accuracy carries the highest weight in MRR calculation. These results validate the effectiveness of TypePro across all variable categories.

\begin{mybox}
    \textbf{Answer of RQ2:} TypePro achieved high accuracy and MRR@5 across elementary, generic, and user-defined, and significantly outperformed all baselines in Top-1 accuracy, showcases strong and more consistent inference accuracy across all variable categories.
\end{mybox}

\subsection{Results of RQ3}
\revise{To answer RQ3, we evaluated the effectiveness of TypePro on state-of-the-art open-source and closed-source large language models. Specifically, we selected two closed-source LLMs (GPT-4.1~\cite{gpt}, claude-4~\cite{sonnet}) and three open-source models (qwen3-coder-480b~\cite{qwen}, llama-4-128x17b~\cite{touvron2023llama}, openPangu-Embedded-7B-V1.1~\cite{chen2025pangu}) to evaluate TypePro. Since TypePro does not require fine-tuning of LLMs, we access all LLMs via API calls. We select the largest open-source models in terms of parameter size in the evaluation.}

\revise{
Table \ref{tab: LLM result} shows the results. We can see that although there are slight differences in the Top-1 EM achieved by TypePro across different large language models, the differences are not significant. The lowest (qwen3-coder-plus) reached 83.8\%. Additionally, we observe that GPT-4.1 performed the best among the five models we selected, achieving a Top-1 EM of 88.9\%. On the other hand, we found that openPangu-Embedded-7B-V1.1 achieved the lowest accuracy of 59.6\%, mainly because the model has fewer parameters. This indicates that TypePro's approach can generalize across different LLMs.}

\begin{table}[h]

\renewcommand{\arraystretch}{1.2} %
    \centering
    \caption{Top-1 Exact Match (\%) accuracy of TypePro in different LLMs. }
    \footnotesize
    \begin{tabular}{l|c|ccc|ccc}
    \toprule
        \textbf{Approach} &\textbf{All} &\textbf{Var} & \textbf{Ret} & \textbf{Arg} & \textbf{Ele} &\textbf{Gen}&\textbf{Usr}  \\
        \toprule
         qwen3-coder-plus~\cite{qwen} & 83.8 &83.3 &87.2 & 83.7 &89.8 &72.5 &84.8 \\
         llama-4-128x17b~\cite{touvron2023llama} & 85.6 & 84.9 &89.2 &86.0&88.6&80.1&85.7\\
        openPangu-Embedded-7B-V1.1~\cite{chen2025pangu} & 59.6 & 61.1 & 63.9 & 52.9 & 50.8 & 65.5 & 74.1\\
         claude-sonnet-4~\cite{sonnet} & 88.7 & 91.2 & 88.1&92.3&82.3&89.2&88.8\\
         GPT-4.1~\cite{gpt}& 88.9& 89.6&88.9&87.2&93.2&82.6&87.6\\
    \bottomrule
    \end{tabular}
    
    \label{tab: LLM result}
\end{table}
\begin{mybox}
\revise{
    \textbf{Answer of RQ3:} TypePro can achieve stable precision in different types of LLMs, showing the generality of TypePro.}
\end{mybox}

\subsection{Results of RQ4}

To answer RQ4, we conducted an ablation study by comparing TypePro with the following baselines: 

\begin{itemize}[leftmargin=*]
\item \textbf{w/o Candidate Type Ranking}: We configured TypePro by removing the candidate type recommendation from the prompt.
\item \textbf{w/o Inter-Procedural Analysis}: We configured TypePro by leveraging the code slicing process of TypeGen, which does not contain sufficient inter-procedural information.
\item \textbf{w/o Slicing}: We configured TypePro by leveraging the contextual information considered by TIGER. Specifically, this context is sampled by taking 300 characters both before and after the position of the target variable.
\item \textbf{w/o All}: We configured TypePro by leveraging the contextual information considered by TIGER, while  removing the candidate type recommendation from the prompt.
\end{itemize}

\begin{table}[h]

\renewcommand{\arraystretch}{1.3} %
    \centering
    \caption{Top-1 Exact Match (\%) accuracy of TypePro when different components of the prompt design in TypePro are removed. }
    \footnotesize
    \begin{tabular}{l|c|ccc|ccc}
    \toprule
        \textbf{Approach} &\textbf{All} &\textbf{Var} & \textbf{Ret} & \textbf{Arg} & \textbf{Ele} &\textbf{Gen}&\textbf{Usr}  \\
        \toprule
         w/o All & 62.0 &67.8 &54.8 & 51.0 &72.5 &56.7 &42.9 \\
         w/o Slicing & 63.9 & 69.2 &58.6 &53.5&72.3&59.2&49.5\\
         w/o Candidate Type Ranking& 83.3 & 82.4 &87.7&84.1 &87.9&77.5&79.3\\
         w/o Inter-Procedural Analysis& 74.7 & 77.7 & 79.7&66.0&80.4&67.0&71.4\\
         \cline{1-8}
         TypePro& \cellcolor{myc2!50}{87.4}& \cellcolor{myc2!50}{87.9}&\cellcolor{myc2!50}{88.6}&\cellcolor{myc2!50}{85.5}&\cellcolor{myc2!50}{91.5}&\cellcolor{myc2!50}{81.7}&\cellcolor{myc2!50}{85.3}\\
    \bottomrule
    \end{tabular}
    
    \label{tab: RQ3 result}
\end{table}

Following the methodology of Peng et al.\cite{peng2023generative}, we compare the performance of TypePro with baseline approaches in terms of Top-1 Exact Match accuracy, as presented in Table \ref{tab: RQ3 result}.
It can be observed that the accuracy of w/o Inter-Procedural Analysis decreases to 74.7\% on the Top-1 Exact Match. Notably, for the Arg and Gen categories, the accuracy drops by 19.5 and 14.7 percentage points, respectively.
Furthermore, when adopting the contextual information extraction process of TIGER (a.k.a., w/o Slicing), the overall accuracy declines by 23.5 percentage points, underscoring the contribution of inter-procedural analysis to provide sufficient contextual information and therefore improve  the accuracy of TypePro.
In addition, removing the candidate type ranking process (a.k.a., w/o Candidate Type Ranking) results in a 3.9 percentage point reduction in overall accuracy, with a particularly impact on the Usr category, where the accuracy decreases by 6.0 percentage points.
These findings demonstrate that candidate type ranking effectively supplements the LLMs with domain knowledge related to user-defined and third-party library data types, thereby enhancing the overall accuracy of TypePro.

\begin{mybox}
    \textbf{Answer of RQ4:} The inter-procedural code slicing in TypePro’s prompt design enhances overall type inference performance by 13.0 percentage points, whereas the candidate type ranking contributes an additional 3.9 percentage point improvement. The above results show the contribution of different components integrated in TypePro.
\end{mybox}

\subsection{Results of RQ5}

To answer RQ5, we run TypePro and the baseline models on the ManyTypes4TypeScript dataset to evaluate the effectiveness of type inference in TypeScript. Specifically, we selected CodeT5, CodeT5+, and UnixCoder as the baselines. CLS approaches (i.e., HiTyper and Type4Py) were excluded mainly because their performance is inferior to the GEN and GEN+SIM approaches, as shown in RQ1. In addition, we did not compare with TypeGen and TIGER, as these approaches are designed specific to Python.  

\begin{table}[t]
\centering
\caption{Comparison with baselines on the ManyTypes4TypeScript dataset}
\tiny
\renewcommand{\arraystretch}{1.3} 
\scalebox{0.98}{
\begin{tabular}{cc|cccc|cccc|cccc|cccc}
\toprule
\multirow{2}{*}{\textbf{Metrics}}& \multirow{2}{*}{\textbf{Approach}} &  \multicolumn{4}{c}{\textbf{Top-1}} &\multicolumn{4}{c}{\textbf{Top-3}} & \multicolumn{4}{c}{\textbf{Top-5}} & \multicolumn{4}{c}{\textbf{MRR@5}}\\
\cline{3-18}
    &                  &   \textbf{Var} & \textbf{Ret} & \textbf{Arg} & \textbf{All} & \textbf{Var} & \textbf{Ret} & \textbf{Arg} & \textbf{All} &\textbf{Var} & \textbf{Ret} & \textbf{Arg} & \textbf{All} &\textbf{Var} & \textbf{Ret} & \textbf{Arg} & \textbf{All} \\
    \toprule
\multirow{4}{*}{\textbf{EM}}
&\multirow{1}{*}{CodeT5} & 75.3 & 85.1 & 69.3 & 74.1 & 85.8 & 89.5 & 78.7 & 83.3 & 87.6 & 91.0 & 80.6 & 85.1& 80.5 & 87.4 & 74.0 & 78.6\\
&CodeT5+ & \cellcolor{myc!50}79.5 & 84.4 & \cellcolor{myc!50}70.5 & \cellcolor{myc!50}76.3 & \cellcolor{myc!50}87.2 &89.9 & \cellcolor{myc!50}78.9 & \cellcolor{myc!50}84.0 & \cellcolor{myc2!50}88.6 &91.5 &80.7 & 85.6& \cellcolor{myc!50}83.4 & 87.3 & \cellcolor{myc!50}74.8 & \cellcolor{myc!50}80.3\\
&UnixCoder & 77.8 & \cellcolor{myc!50}85.5 & 68.4 & 74.9 & 86.5 &\cellcolor{myc!50} 90.3&78.8&83.7 &\cellcolor{myc2!50}88.6& \cellcolor{myc!50}91.8 & 81.2 & \cellcolor{myc!50}85.9& 82.2 & \cellcolor{myc!50}88.0 & 73.7 & 79.4\\ 
& \multirow{1}{*}{TypePro} & \cellcolor{myc2!50}85.6 & \cellcolor{myc2!50}89.1 & \cellcolor{myc2!50}86.7 & \cellcolor{myc2!50}86.6 & \cellcolor{myc2!50}88.5 & \cellcolor{myc2!50}92.1 & \cellcolor{myc2!50}91.1 & \cellcolor{myc2!50}90.1 & \cellcolor{myc!50}88.5 & \cellcolor{myc2!50}92.3 & \cellcolor{myc2!50}91.2 &\cellcolor{myc2!50}90.2 & \cellcolor{myc2!50}87.0 & \cellcolor{myc2!50}90.6 & \cellcolor{myc2!50}88.9 & \cellcolor{myc2!50}88.3\\

\midrule
\multirow{4}{*}{\textbf{BM}}
&\multirow{1}{*}{CodeT5} & 75.8 & 85.5 & 69.6 & 74.4 & 86.0 & 89.7 & 79.0 & 83.5 & 88.0 & 91.3 & 81.0 & 85.5& 80.9 & 87.8 & 74.3 & 79.0\\
&CodeT5+ & \cellcolor{myc!50}79.6 & 84.7 & \cellcolor{myc!50}70.7 & \cellcolor{myc!50}76.5 & \cellcolor{myc!50}87.6 &\cellcolor{myc!50}90.5 & \cellcolor{myc!50}79.5 & \cellcolor{myc!50}84.6 & \cellcolor{myc2!50}89.2 &\cellcolor{myc!50}92.3 &81.4 & \cellcolor{myc!50}86.3& \cellcolor{myc!50}83.7 & 87.7 & \cellcolor{myc!50}75.1 & \cellcolor{myc!50}80.6\\
&UnixCoder & 77.9 & \cellcolor{myc!50}85.8 & 68.6 & 75.0 & 86.8 & 90.4&79.2&84.0 &\cellcolor{myc!50}88.9& 92.0 & \cellcolor{myc!50}81.7 & \cellcolor{myc!50}86.3& 82.4 & \cellcolor{myc!50}88.2&74.0& 79.6\\
& \multirow{1}{*}{TypePro} & \cellcolor{myc2!50}85.7 & \cellcolor{myc2!50}89.3 & \cellcolor{myc2!50}86.7 & \cellcolor{myc2!50}86.6 & \cellcolor{myc2!50}88.5 & \cellcolor{myc2!50}92.2 & \cellcolor{myc2!50}91.1 & \cellcolor{myc2!50}90.2 & 88.6 & \cellcolor{myc2!50}92.4 & \cellcolor{myc2!50}92.4 &\cellcolor{myc2!50}90.3&\cellcolor{myc2!50}87.1 & \cellcolor{myc2!50}90.8 & \cellcolor{myc2!50}88.9 & \cellcolor{myc2!50}88.4  \\

\bottomrule
\end{tabular}
}
\label{tab: RQ4 result}
\end{table}

Table \ref{tab: RQ4 result} presents the results. As we can see, TypePro demonstrates comprehensive outperformance over all baseline models on both the Exact Match and Base Match metrics, with only a marginal shortfall observed in the Top-5 Var category. Notably, TypePro surpasses the best baseline (CodeT5+) by 10.3\% in Top-1 Exact Match, and achieves an accuracy of 85.6\% for the Var category, 89.1\% for the Ret category, and 86.7\% for the Arg category in Top-1 Exact Match.
On the other hand, the improvement in the Ret category for ManyTypes4TypeScript is smaller compared to ManyTypes4Py. This is mainly because more than one-third of Ret in the TypeScript dataset are labeled as \texttt{void}, and function bodies are typically short, allowing CLS-based models to capture the complete context within their input window. In contrast, for the Arg category, TypePro achieves an improvement of approximately 20\% over the baselines. These results demonstrate that TypePro exhibits robustness across different programming languages.

\begin{mybox}
    \textbf{Answer of RQ5:} TypePro outperforms all baselines on both Exact
Match and Base Match metrics in ManyTypes4TypeScript, showing the effectiveness of robustness of TypePro in different programming languages.
\end{mybox}

\section{Limitations}

    \subsection{ Limitations of the LLM-Based Method} In our experiments, we used a frequency-based voting mechanism on multiple sampling outputs to generate Top-N results from LLMs, instead of directly ranking candidates by generative likelihood as in CodeT5. The quality of this approach essentially depends on two factors: the diversity of the LLM API responses and the configuration of API parameters (e.g., temperature). Specifically, if the temperature is set too low, the model tends to generate identical or very similar outputs across multiple samplings, increasing the frequency of certain results and reducing the diversity of the Top-N outputs. Conversely, if we set a higher temperature for the LLM, although this increases output diversity, as mentioned earlier, in some cases the candidate type ranking process cannot consistently provide a sufficient number of candidate types due to threshold limitations. In such cases, relying solely on the LLM for semantic type inference leads to limited improvement in Top-3 and Top-5 accuracy of TypeGen compared to baseline methods. In the future, the candidate type ranking process can enhance the sampling around the target variable and provide richer syntactic and semantic guidance for LLMs. 

    \subsection{Limitations in the Experimental Setup} Due to the unavailability of DLInfer's source code~\cite{yan2023dlinfer}, we were unable to include it in our evaluation. Although both DLInfer and our method use static code slicing, we chose to compare our method with TypeGen, which also combines code slicing with LLMs. Nevertheless, this limitation does not affect the validity of our comparative analysis with other baseline methods.

\section{Conclusion}
This paper introduces TypePro, a novel type inference method for dynamic languages based on code slicing and LLMs. TypePro incorporates inter-procedural information related to variables into static code slicing and traces variables with dependency relationships. It then recommends candidate types by calculating text similarity between the initially generated results from the LLMs and variable declaration-related information. Extensive experiments on the ManyTypes4Py and ManyTypes4TypeScript datasets demonstrate that TypePro achieves  improvements across various type categories outperforming existing Python baselines in different programming languages. Noteably, TypePro achieves the highest Top-1 accuracy comparing to the baselines, showing the usefulness in the scenarios that require full automation (e.g., compilation speedup). 

\section{Data Availability}

We released the implementation of TypePro, together with the raw data of evaluation results (RQ1--5), on the project website \url{https://github.com/umxadmin/TypePro}. 

\section*{Acknowledgment}

This work was supported by the National Natural Science Foundation of China (Grant No. 62402405), Youth Program of the Xiamen Natural Science Foundation (Grant No. 3502Z202471016), and the Fundamental Research Funds for the Central Universities (Grant No. 20720240087). Huaxun Huang is the corresponding author.

\bibliographystyle{ACM-Reference-Format}
\bibliography{reference}

@inproceedings{Typilus,
author = {Allamanis, Miltiadis and Barr, Earl T. and Ducousso, Soline and Gao, Zheng},
title = {Typilus: neural type hints},
year = {2020},
isbn = {9781450376136},
publisher = {Association for Computing Machinery},
address = {New York, NY, USA},
url = {https://doi.org/10.1145/3385412.3385997},
doi = {10.1145/3385412.3385997},
booktitle = {Proceedings of the 41st ACM SIGPLAN Conference on Programming Language Design and Implementation},
pages = {91–105},
numpages = {15},
location = {London, UK},
series = {PLDI 2020}
}

@inproceedings{parnin2011automated,
  title={Are automated debugging techniques actually helping programmers?},
  author={Parnin, Chris and Orso, Alessandro},
  booktitle={Proceedings of the 2011 international symposium on software testing and analysis},
  pages={199--209},
  year={2011}
}

@misc{pep484,
  author = {Guido van Rossum and Jukka Lehtosalo and Łukasz Langa},
  title = {{PEP 484 – Type Hints}},
  year = {2014},
  howpublished = {[Online]. Available: \url{https://peps.python.org/pep-0484/}},
  note = {Accessed: 2025-08-27}
}

@inproceedings{bierman2014understanding,
  title={Understanding typescript},
  author={Bierman, Gavin and Abadi, Mart{\'\i}n and Torgersen, Mads},
  booktitle={European Conference on Object-Oriented Programming},
  pages={257--281},
  year={2014},
  organization={Springer},
}

@inproceedings{hellendoorn2018deep,
  title={Deep learning type inference},
  author={Hellendoorn, Vincent J and Bird, Christian and Barr, Earl T and Allamanis, Miltiadis},
  booktitle={Proceedings of the 2018 26th acm joint meeting on european software engineering conference and symposium on the foundations of software engineering},
  pages={152--162},
  year={2018},
}

@inproceedings{DoMachineLearning,
  doi = {10.4230/LIPICS.ECOOP.2023.37},
  url = {https://drops.dagstuhl.de/entities/document/10.4230/LIPIcs.ECOOP.2023.37},
  
  author = {Yee, Ming-Ho and Guha, Arjun},
  
  language = {en},
  
  title = {Do Machine Learning Models Produce TypeScript Types That Type Check?},
  
  publisher = {Schloss Dagstuhl – Leibniz-Zentrum für Informatik},
  
  year = {2023},
  
  copyright = {Creative Commons Attribution 4.0 International license}
}

@online{mypy,
  title     = {Mypy: Static Typing for Python},
  year      = {2025},
  url       = {https://github.com/python/mypy/},
  urldate   = {2025-09-08}
}

@online{sonnet,
  title     = {Claude Sonnet 4.6},
  year      = {2025},
  url       = {https://www.anthropic.com/claude/sonnet},
  urldate   = {2025-09-08}
}

@online{gpt,
  title     = {ChatGPT},
  year      = {2025},
  url       = {https://chat.chatbot.app/gpt4},
  urldate   = {2025-09-08}
}

@article{touvron2023llama,
  title={Llama: Open and efficient foundation language models. arXiv 2023},
  author={Touvron, Hugo and Lavril, Thibaut and Izacard, Gautier and Martinet, Xavier and Lachaux, Marie-Anne and Lacroix, Timoth{\'e}e and Rozi{\`e}re, Baptiste and Goyal, Naman and Hambro, Eric and Azhar, Faisal and others},
  journal={arXiv preprint arXiv:2302.13971},
  volume={10},
  year={2023}
}

@online{qwen,
  title     = {Qwen Chat},
  year      = {2025},
  url       = {https://chat.qwen.ai/},
  urldate   = {2025-09-08}
}

@inproceedings{yan2023dlinfer,
  title={Dlinfer: Deep learning with static slicing for python type inference},
  author={Yan, Yanyan and Feng, Yang and Fan, Hongcheng and Xu, Baowen},
  booktitle={2023 IEEE/ACM 45th International Conference on Software Engineering (ICSE)},
  pages={2009--2021},
  year={2023},
  organization={IEEE}
}

@article{chen2025pangu,
  title={Pangu embedded: An efficient dual-system llm reasoner with metacognition},
  author={Chen, Hanting and Wang, Yasheng and Han, Kai and Li, Dong and Li, Lin and Bi, Zhenni and Li, Jinpeng and Wang, Haoyu and Mi, Fei and Zhu, Mingjian and others},
  journal={arXiv preprint arXiv:2505.22375},
  year={2025}
}

@inproceedings{salis2021pycg,
  title={Pycg: Practical call graph generation in python},
  author={Salis, Vitalis and Sotiropoulos, Thodoris and Louridas, Panos and Spinellis, Diomidis and Mitropoulos, Dimitris},
  booktitle={2021 IEEE/ACM 43rd International Conference on Software Engineering (ICSE)},
  pages={1646--1657},
  year={2021},
  organization={IEEE}
}

@inproceedings{staicu2018understanding,
  title={Understanding and automatically preventing injection attacks on Node. js},
  author={Staicu, Cristian-Alexandru and Pradel, Michael and Livshits, Ben},
  booktitle={Network and Distributed System Security Symposium (NDSS)},
  year={2018}
}

@inproceedings{guarnieri2009gatekeeper,
  title={GATEKEEPER: Mostly Static Enforcement of Security and Reliability Policies for JavaScript Code.},
  author={Guarnieri, Salvatore and Livshits, V Benjamin},
  booktitle={USENIX Security Symposium},
  volume={10},
  pages={78--85},
  year={2009}
}

@article{horwitz1990interprocedural,
  title={Interprocedural slicing using dependence graphs},
  author={Horwitz, Susan and Reps, Thomas and Binkley, David},
  journal={ACM Transactions on Programming Languages and Systems (TOPLAS)},
  volume={12},
  number={1},
  pages={26--60},
  year={1990},
  publisher={ACM New York, NY, USA}
}

@inproceedings{mir2022type4py,
  title={Type4py: Practical deep similarity learning-based type inference for python},
  author={Mir, Amir M and Lato{\v{s}}kinas, Evaldas and Proksch, Sebastian and Gousios, Georgios},
  booktitle={Proceedings of the 44th International Conference on Software Engineering},
  pages={2241--2252},
  year={2022}
}

@inproceedings{peng2023generative,
  title={Generative type inference for python},
  author={Peng, Yun and Wang, Chaozheng and Wang, Wenxuan and Gao, Cuiyun and Lyu, Michael R},
  booktitle={2023 38th IEEE/ACM International Conference on Automated Software Engineering (ASE)},
  pages={988--999},
  year={2023},
  organization={IEEE}
}

@article{wang2024tiger,
  title={Tiger: A generating-then-ranking framework for practical python type inference},
  author={Wang, Chong and Zhang, Jian and Lou, Yiling and Liu, Mingwei and Sun, Weisong and Liu, Yang and Peng, Xin},
  journal={arXiv preprint arXiv:2407.02095},
  year={2024}
}

@inproceedings{mir2021manytypes4py,
  title={Manytypes4py: A benchmark python dataset for machine learning-based type inference},
  author={Mir, Amir M and Lato{\v{s}}kinas, Evaldas and Gousios, Georgios},
  booktitle={2021 IEEE/ACM 18th International Conference on Mining Software Repositories (MSR)},
  pages={585--589},
  year={2021},
  organization={IEEE}
}

@inproceedings{Manytypes4typescript,
author = {Jesse, Kevin and Devanbu, Premkumar T.},
title = {ManyTypes4TypeScript: a comprehensive TypeScript dataset for sequence-based type inference},
year = {2022},
isbn = {9781450393034},
publisher = {Association for Computing Machinery},
address = {New York, NY, USA},
url = {https://doi.org/10.1145/3524842.3528507},
doi = {10.1145/3524842.3528507},
booktitle = {Proceedings of the 19th International Conference on Mining Software Repositories},
pages = {294–298},
numpages = {5},
location = {Pittsburgh, Pennsylvania},
series = {MSR '22}
}

@inproceedings{xu2016python,
  title={Python probabilistic type inference with natural language support},
  author={Xu, Zhaogui and Zhang, Xiangyu and Chen, Lin and Pei, Kexin and Xu, Baowen},
  booktitle={Proceedings of the 2016 24th ACM SIGSOFT international symposium on foundations of software engineering},
  pages={607--618},
  year={2016}
}

@software{microsoft_pyright,
  author = {{Microsoft}},
  title = {Pyright - Static Type Checker for Python},
  year = {2023},
  url = {https://github.com/microsoft/pyright},
  urldate = {2025-08-27}
}

@software{google_pytype,
  author = {{Google}},
  title = {pytype - A Static Type Analyzer for Python Code},
  year = {2023},
  url = {https://github.com/google/pytype},
  urldate = {2025-08-27}
}

@article{DataFlowInference,
author = {Pavlinovic, Zvonimir and Su, Yusen and Wies, Thomas},
title = {Data flow refinement type inference},
year = {2021},
issue_date = {January 2021},
publisher = {Association for Computing Machinery},
address = {New York, NY, USA},
volume = {5},
number = {POPL},
url = {https://doi.org/10.1145/3434300},
doi = {10.1145/3434300},
journal = {Proc. ACM Program. Lang.},
month = jan,
articleno = {19},
numpages = {31}
}

@inproceedings{Symbolicabstract,
author = {Emmi, Michael and Enea, Constantin},
title = {Symbolic abstract data type inference},
year = {2016},
isbn = {9781450335492},
publisher = {Association for Computing Machinery},
address = {New York, NY, USA},
url = {https://doi.org/10.1145/2837614.2837645},
doi = {10.1145/2837614.2837645},
booktitle = {Proceedings of the 43rd Annual ACM SIGPLAN-SIGACT Symposium on Principles of Programming Languages},
pages = {513–525},
numpages = {13},
location = {St. Petersburg, FL, USA},
series = {POPL '16}
}

@misc{wei2020lambdanetprobabilistictypeinference,
      title={LambdaNet: Probabilistic Type Inference using Graph Neural Networks}, 
      author={Jiayi Wei and Maruth Goyal and Greg Durrett and Isil Dillig},
      year={2020},
      eprint={2005.02161},
      archivePrefix={arXiv},
      primaryClass={cs.PL},
      url={https://arxiv.org/abs/2005.02161}, 
}

@misc{guo2022unixcoderunifiedcrossmodalpretraining,
      title={UniXcoder: Unified Cross-Modal Pre-training for Code Representation}, 
      author={Daya Guo and Shuai Lu and Nan Duan and Yanlin Wang and Ming Zhou and Jian Yin},
      year={2022},
      eprint={2203.03850},
      archivePrefix={arXiv},
      primaryClass={cs.CL},
      url={https://arxiv.org/abs/2203.03850}, 
}

@misc{fried2023incodergenerativemodelcode,
      title={InCoder: A Generative Model for Code Infilling and Synthesis}, 
      author={Daniel Fried and Armen Aghajanyan and Jessy Lin and Sida Wang and Eric Wallace and Freda Shi and Ruiqi Zhong and Wen-tau Yih and Luke Zettlemoyer and Mike Lewis},
      year={2023},
      eprint={2204.05999},
      archivePrefix={arXiv},
      primaryClass={cs.SE},
      url={https://arxiv.org/abs/2204.05999}, 
}

@article{BM25,
url = {http://dx.doi.org/10.1561/1500000019},
year = {2009},
volume = {3},
journal = {Foundations and Trends® in Information Retrieval},
title = {The Probabilistic Relevance Framework: BM25 and Beyond},
doi = {10.1561/1500000019},
issn = {1554-0669},
number = {4},
pages = {333-389},
author = {Stephen Robertson and Hugo Zaragoza}
}

@article{codeT5,
  title={Codet5: Identifier-aware unified pre-trained encoder-decoder models for code understanding and generation},
  author={Wang, Yue and Wang, Weishi and Joty, Shafiq and Hoi, Steven CH},
  journal={arXiv preprint arXiv:2109.00859},
  year={2021}
}

@inproceedings{Hityper,
  title={Static inference meets deep learning: a hybrid type inference approach for python},
  author={Peng, Yun and Gao, Cuiyun and Li, Zongjie and Gao, Bowei and Lo, David and Zhang, Qirun and Lyu, Michael},
  booktitle={Proceedings of the 44th International Conference on Software Engineering},
  pages={2019--2030},
  year={2022}
}

@article{codeT5+,
  title={Codet5+: Open code large language models for code understanding and generation},
  author={Wang, Yue and Le, Hung and Gotmare, Akhilesh Deepak and Bui, Nghi DQ and Li, Junnan and Hoi, Steven CH},
  journal={arXiv preprint arXiv:2305.07922},
  year={2023}
}

@software{huggingface2025hub,
  author = {{Hugging Face Inc.}},
  title = {{Hugging Face Hub: A platform for sharing machine learning models, datasets and demos}},
  year = {2025},
  url = {https://huggingface.co/},
  urldate = {2025-09-05}
}

@article{howFar,
author = {Guo, Yimeng and Chen, Zhifei and Chen, Lin and Xu, Wenjie and Li, Yanhui and Zhou, Yuming and Xu, Baowen},
title = {Generating Python Type Annotations from Type Inference: How Far Are We?},
year = {2024},
issue_date = {June 2024},
publisher = {Association for Computing Machinery},
address = {New York, NY, USA},
volume = {33},
number = {5},
issn = {1049-331X},
url = {https://doi.org/10.1145/3652153},
doi = {10.1145/3652153},
journal = {ACM Trans. Softw. Eng. Methodol.},
month = jun,
articleno = {123},
numpages = {38}
}

@book{MRRs,
  title     = {Introduction to Information Retrieval},
  author    = {Manning, Christopher D. and Schütze, Hinrich and Raghavan, Prabhakar},
  year      = {2008},
  publisher = {Cambridge University Press},
  address   = {Cambridge, UK}
}

\end{document}